\pdfoutput=1

\documentclass[12pt,a4paper]{article}

\usepackage{ifthen} 
\newboolean{pdflatex}
\setboolean{pdflatex}{true} 

\newboolean{articletitles}
\setboolean{articletitles}{true} 

\newboolean{uprightparticles}
\setboolean{uprightparticles}{false} 

\newboolean{inbibliography}
\setboolean{inbibliography}{false} 

\def\paperauthors{LHCb collaboration} 
\def\paperasciititle{Search for CP violation using triple product asymmetries in Lb->pK-pi+pi-, Lb->pK-K+K- and Xib0->pK-K-pi+ decays
} 
\def\papertitle{Search for \CP violation using triple product asymmetries in \LbTopKpipi, \LbTopKKK and \XibTopKKpiWS decays} 

\def\paperkeywords{{High Energy Physics}, {LHCb}} 
\def\papercopyright{\the\year\ CERN for the benefit of the LHCb collaboration} 
\def\paperlicence{CC-BY-4.0 licence}
\def\paperlicenceurl{https://creativecommons.org/licenses/by/4.0/}

\usepackage[top=1in, bottom=1.25in, left=1in, right=1in]{geometry}

\usepackage{microtype}
\usepackage{lineno}  
\usepackage{xspace} 
\usepackage{caption} 

\usepackage{graphicx}  
\usepackage{color}
\usepackage{colortbl}
\graphicspath{{./figs/}} 

\usepackage{amsmath} 
\usepackage{amssymb}
\usepackage{amsfonts}
\usepackage{upgreek} 

\newcommand*\patchAmsMathEnvironmentForLineno[1]{%
\expandafter\let\csname old#1\expandafter\endcsname\csname #1\endcsname
\expandafter\let\csname oldend#1\expandafter\endcsname\csname
end#1\endcsname
 \renewenvironment{#1}%
   {\linenomath\csname old#1\endcsname}%
   {\csname oldend#1\endcsname\endlinenomath}%
}
\newcommand*\patchBothAmsMathEnvironmentsForLineno[1]{%
  \patchAmsMathEnvironmentForLineno{#1}%
  \patchAmsMathEnvironmentForLineno{#1*}%
}
\AtBeginDocument{%
\patchBothAmsMathEnvironmentsForLineno{equation}%
\patchBothAmsMathEnvironmentsForLineno{align}%
\patchBothAmsMathEnvironmentsForLineno{flalign}%
\patchBothAmsMathEnvironmentsForLineno{alignat}%
\patchBothAmsMathEnvironmentsForLineno{gather}%
\patchBothAmsMathEnvironmentsForLineno{multline}%
\patchBothAmsMathEnvironmentsForLineno{eqnarray}%
}

\usepackage{hyperxmp}

\usepackage[pdftex,
            pdfauthor={\paperauthors},
            pdftitle={\paperasciititle},
            pdfkeywords={\paperkeywords},
            pdfcopyright={Copyright (C) \papercopyright},
            pdflicenseurl={\paperlicenceurl}]{hyperref}

\usepackage[all]{hypcap} 

\usepackage{xspace} 
\usepackage{upgreek}

\def\lhcb {\mbox{LHCb}\xspace}

\def\MagUp {\mbox{\em Mag\kern -0.05em Up}\xspace}

\ifthenelse{\boolean{uprightparticles}}%
{

 \def\Pmu         {\ensuremath{\upmu}\xspace}

 \def\Ppi         {\ensuremath{\uppi}\xspace}

 \def\Pphi        {\ensuremath{\upphi}\xspace}

 \def\Ppsi        {\ensuremath{\uppsi}\xspace}

 \def\PDelta      {\ensuremath{\Delta}\xspace}                 
 \def\PXi      {\ensuremath{\Xi}\xspace}                 
 \def\PLambda      {\ensuremath{\Lambda}\xspace}                 
 \def\PSigma      {\ensuremath{\Sigma}\xspace}                 
 \def\POmega      {\ensuremath{\Omega}\xspace}                 
 \def\PUpsilon      {\ensuremath{\Upsilon}\xspace}                 
 
 \def\PB      {\ensuremath{\mathrm{B}}\xspace}                 
                  
 \def\PD      {\ensuremath{\mathrm{D}}\xspace}

 \def\PJ      {\ensuremath{\mathrm{J}}\xspace}                 
 \def\PK      {\ensuremath{\mathrm{K}}\xspace}

 \def\PX      {\ensuremath{\mathrm{X}}\xspace}

 \def\Pb      {\ensuremath{\mathrm{b}}\xspace}                 
 \def\Pc      {\ensuremath{\mathrm{c}}\xspace}

 \def\Pi      {\ensuremath{\mathrm{i}}\xspace}

 \def\Pp      {\ensuremath{\mathrm{p}}\xspace}

 \def\Ps      {\ensuremath{\mathrm{s}}\xspace}                 
                  
 \def\Pu      {\ensuremath{\mathrm{u}}\xspace}

}
{

 \def\Pmu         {\ensuremath{\mu}\xspace}

 \def\Ppi         {\ensuremath{\pi}\xspace}

 \def\Pphi        {\ensuremath{\phi}\xspace}

 \def\Ppsi        {\ensuremath{\psi}\xspace}                 
                  
 \mathchardef\PDelta="7101
 \mathchardef\PXi="7104
 \mathchardef\PLambda="7103
 \mathchardef\PSigma="7106
 \mathchardef\POmega="710A
 \mathchardef\PUpsilon="7107
                  
 \def\PB      {\ensuremath{B}\xspace}                 
                  
 \def\PD      {\ensuremath{D}\xspace}

 \def\PJ      {\ensuremath{J}\xspace}                 
 \def\PK      {\ensuremath{K}\xspace}

 \def\PX      {\ensuremath{X}\xspace}

 \def\Pb      {\ensuremath{b}\xspace}                 
 \def\Pc      {\ensuremath{c}\xspace}

 \def\Pi      {\ensuremath{i}\xspace}

 \def\Pp      {\ensuremath{p}\xspace}

 \def\Ps      {\ensuremath{s}\xspace}                 
                  
 \def\Pu      {\ensuremath{u}\xspace}

}

\makeatletter
\ifcase \@ptsize \relax
  \newcommand{\miniscule}{\@setfontsize\miniscule{4}{5}}
\or
  \newcommand{\miniscule}{\@setfontsize\miniscule{5}{6}}
\or
  \newcommand{\miniscule}{\@setfontsize\miniscule{5}{6}}
\fi
\makeatother

\DeclareRobustCommand{\optbar}[1]{\shortstack{{\miniscule (\rule[.5ex]{1.25em}{.18mm})}
  \\ [-.7ex] $#1$}}

\def\mumu       {{\ensuremath{\Pmu^+\Pmu^-}}\xspace}

\def\uquark    {{\ensuremath{\Pu}}\xspace}
\def\uquarkbar {{\ensuremath{\overline \uquark}}\xspace}

\def\squark    {{\ensuremath{\Ps}}\xspace}

\def\cquark    {{\ensuremath{\Pc}}\xspace}

\def\bquark    {{\ensuremath{\Pb}}\xspace}

\def\pion   {{\ensuremath{\Ppi}}\xspace}
\def\piz    {{\ensuremath{\pion^0}}\xspace}

\def\pip    {{\ensuremath{\pion^+}}\xspace}
\def\pim    {{\ensuremath{\pion^-}}\xspace}

\def\kaon    {{\ensuremath{\PK}}\xspace}
  \def\Kbar    {{\kern 0.2em\overline{\kern -0.2em \PK}{}}\xspace}

\def\KorKbar    {\kern 0.18em\optbar{\kern -0.18em K}{}\xspace}

\def\Kp      {{\ensuremath{\kaon^+}}\xspace}
\def\Km      {{\ensuremath{\kaon^-}}\xspace}

\def\Kstarz  {{\ensuremath{\kaon^{*}(892)^{0}}}\xspace}
\def\Kstarzb {{\ensuremath{\Kbar{}^{*}(892)^{0}}}\xspace}

\def\Kstarm  {{\ensuremath{\kaon^{*-}}}\xspace}

\newcommand{\phiz}{\ensuremath{\Pphi}\xspace}

  \def\Dbar    {{\kern 0.2em\overline{\kern -0.2em \PD}{}}\xspace}
\def\D       {{\ensuremath{\PD}}\xspace}

\def\DorDbar    {\kern 0.18em\optbar{\kern -0.18em D}{}\xspace}
\def\Dz      {{\ensuremath{\D^0}}\xspace}

\def\B       {{\ensuremath{\PB}}\xspace}
\def\Bbar    {{\ensuremath{\kern 0.18em\overline{\kern -0.18em \PB}{}}}\xspace}

\def\BorBbar    {\kern 0.18em\optbar{\kern -0.18em B}{}\xspace}
\def\Bz      {{\ensuremath{\B^0}}\xspace}

\def\Bs      {{\ensuremath{\B^0_\squark}}\xspace}

\def\jpsi     {{\ensuremath{{\PJ\mskip -3mu/\mskip -2mu\Ppsi\mskip 2mu}}}\xspace}

  \def\Y#1S{\ensuremath{\PUpsilon{(#1S)}}\xspace}

\def\proton      {{\ensuremath{\Pp}}\xspace}

\def\Xires       {{\ensuremath{\PXi}}\xspace}
\def\Xiresbar    {{\ensuremath{\overline \Xires}}\xspace}
\def\Lz          {{\ensuremath{\PLambda}}\xspace}
\def\Lbar        {{\ensuremath{\kern 0.1em\overline{\kern -0.1em\PLambda}}}\xspace}
\def\LorLbar    {\kern 0.18em\optbar{\kern -0.18em \PLambda}{}\xspace}

\def\Lb      {{\ensuremath{\Lz^0_\bquark}}\xspace}
\def\Lbbar   {{\ensuremath{\Lbar{}^0_\bquark}}\xspace}
\def\Lc      {{\ensuremath{\Lz^+_\cquark}}\xspace}

\def\Xibz    {{\ensuremath{\Xires^0_\bquark}}\xspace}

\def\Xibbarz {{\ensuremath{\Xiresbar{}_\bquark^0}}\xspace}

\def\Xb          {{\ensuremath{\PX^0_\bquark}}\xspace}
\def\Xbbar     {{\ensuremath{   \kern 0.2em\overline{\kern -0.2em\PX}  ^0_\bquark}}\xspace}

\newcommand{\decay}[2]{\ensuremath{#1\!\to #2}\xspace}         

\def\to                 {\ensuremath{\rightarrow}\xspace}

\def\CP                {{\ensuremath{C\!P}}\xspace}

\def\AT#1     {\ensuremath{A_{\mathrm{T}}^{#1}}\xspace}           

\def\C#1      {\ensuremath{\mathcal{C}_{#1}}\xspace}                       
\def\Cp#1     {\ensuremath{\mathcal{C}_{#1}^{'}}\xspace}                    
\def\Ceff#1   {\ensuremath{\mathcal{C}_{#1}^{\mathrm{(eff)}}}\xspace}        
\def\Cpeff#1  {\ensuremath{\mathcal{C}_{#1}^{'\mathrm{(eff)}}}\xspace}       
\def\Ope#1    {\ensuremath{\mathcal{O}_{#1}}\xspace}                       
\def\Opep#1   {\ensuremath{\mathcal{O}_{#1}^{'}}\xspace}                    

\newcommand{\tev}{\ifthenelse{\boolean{inbibliography}}{\ensuremath{~T\kern -0.05em eV}}{\ensuremath{\mathrm{\,Te\kern -0.1em V}}}\xspace}
\newcommand{\gev}{\ensuremath{\mathrm{\,Ge\kern -0.1em V}}\xspace}
\newcommand{\mev}{\ensuremath{\mathrm{\,Me\kern -0.1em V}}\xspace}
\newcommand{\kev}{\ensuremath{\mathrm{\,ke\kern -0.1em V}}\xspace}
\newcommand{\ev}{\ensuremath{\mathrm{\,e\kern -0.1em V}}\xspace}
\newcommand{\gevc}{\ensuremath{{\mathrm{\,Ge\kern -0.1em V\!/}c}}\xspace}
\newcommand{\mevc}{\ensuremath{{\mathrm{\,Me\kern -0.1em V\!/}c}}\xspace}
\newcommand{\gevcc}{\ensuremath{{\mathrm{\,Ge\kern -0.1em V\!/}c^2}}\xspace}
\newcommand{\gevgevcccc}{\ensuremath{{\mathrm{\,Ge\kern -0.1em V^2\!/}c^4}}\xspace}
\newcommand{\mevcc}{\ensuremath{{\mathrm{\,Me\kern -0.1em V\!/}c^2}}\xspace}

\def\mum  {\ensuremath{{\,\upmu\mathrm{m}}}\xspace}

\def\invfb   {\ensuremath{\mbox{\,fb}^{-1}}\xspace}

\newcommand{\chisq}{\ensuremath{\chi^2}\xspace}

\newcommand{\chisqip}{\ensuremath{\chi^2_{\text{IP}}}\xspace}

\def\gsim{{~\raise.15em\hbox{$>$}\kern-.85em
          \lower.35em\hbox{$\sim$}~}\xspace}
\def\lsim{{~\raise.15em\hbox{$<$}\kern-.85em
          \lower.35em\hbox{$\sim$}~}\xspace}

\def\sPlot{\mbox{\em sPlot}\xspace}

\def\ptot       {\mbox{$p$}\xspace}
\def\pt         {\mbox{$p_{\mathrm{ T}}$}\xspace}

\def\evtgen     {\mbox{\textsc{EvtGen}}\xspace}

\def\geant      {\mbox{\textsc{Geant4}}\xspace}

\def\photos     {\mbox{\textsc{Photos}}\xspace}

\def\pythia     {\mbox{\textsc{Pythia}}\xspace}

\def\tell1  {TELL1\xspace}
\def\ukl1   {UKL1\xspace}

\usepackage{cite} 
\usepackage{mciteplus}

\usepackage[normalem]{ulem}
\usepackage{xcolor}
\newcommand\redsout{\bgroup\markoverwith{\textcolor{red}{\rule[0.5ex]{2pt}{0.4pt}}}\ULon}

\usepackage{booktabs}
\usepackage{multirow}

\def\LbToppipipi {\decay{\Lb}{\proton\pim\pip\pim}}
\def\LbTopKpipi  {\decay{\Lb}{\proton\Km\pip\pim}}
\def\LbTopKKK    {\decay{\Lb}{\proton\Km\Kp\Km}}
\def\XibTopKKpiWS{\decay{\Xibz}{\proton\Km\Km\pip}}
\def\That            {\ensuremath{\widehat{T}}\xspace}
\newcommand{\At}     {{\ensuremath{A_{\widehat{T}}}}\xspace}
\newcommand{\Atbar}  {{\ensuremath{\kern 0.2em\overline{\kern -0.2em A}_{\widehat{T}}}}\xspace}
\newcommand{\aCPTodd}{{\ensuremath{a_\CP^{\widehat{T}\text{-odd}}}}\xspace}
\newcommand{\aPTodd} {{\ensuremath{a_P^{\widehat{T}\text{-odd}}}}\xspace}
\newcommand{\CT}     {{\ensuremath{C_{\widehat{T}}}}\xspace}
\newcommand{\CTbar}  {{\ensuremath{\kern 0.1em\overline{\kern -0.1em C}_{\widehat{T}}}}\xspace}

\usepackage{longtable} 

\begin{document}

\renewcommand{\thefootnote}{\fnsymbol{footnote}}
\setcounter{footnote}{1}


\begin{titlepage}
\pagenumbering{roman}

\vspace*{-1.5cm}
\centerline{\large EUROPEAN ORGANIZATION FOR NUCLEAR RESEARCH (CERN)}
\vspace*{1.5cm}
\noindent
\begin{tabular*}{\linewidth}{lc@{\extracolsep{\fill}}r@{\extracolsep{0pt}}}
\ifthenelse{\boolean{pdflatex}}
{\vspace*{-1.5cm}\mbox{\!\!\!\includegraphics[width=.14\textwidth]{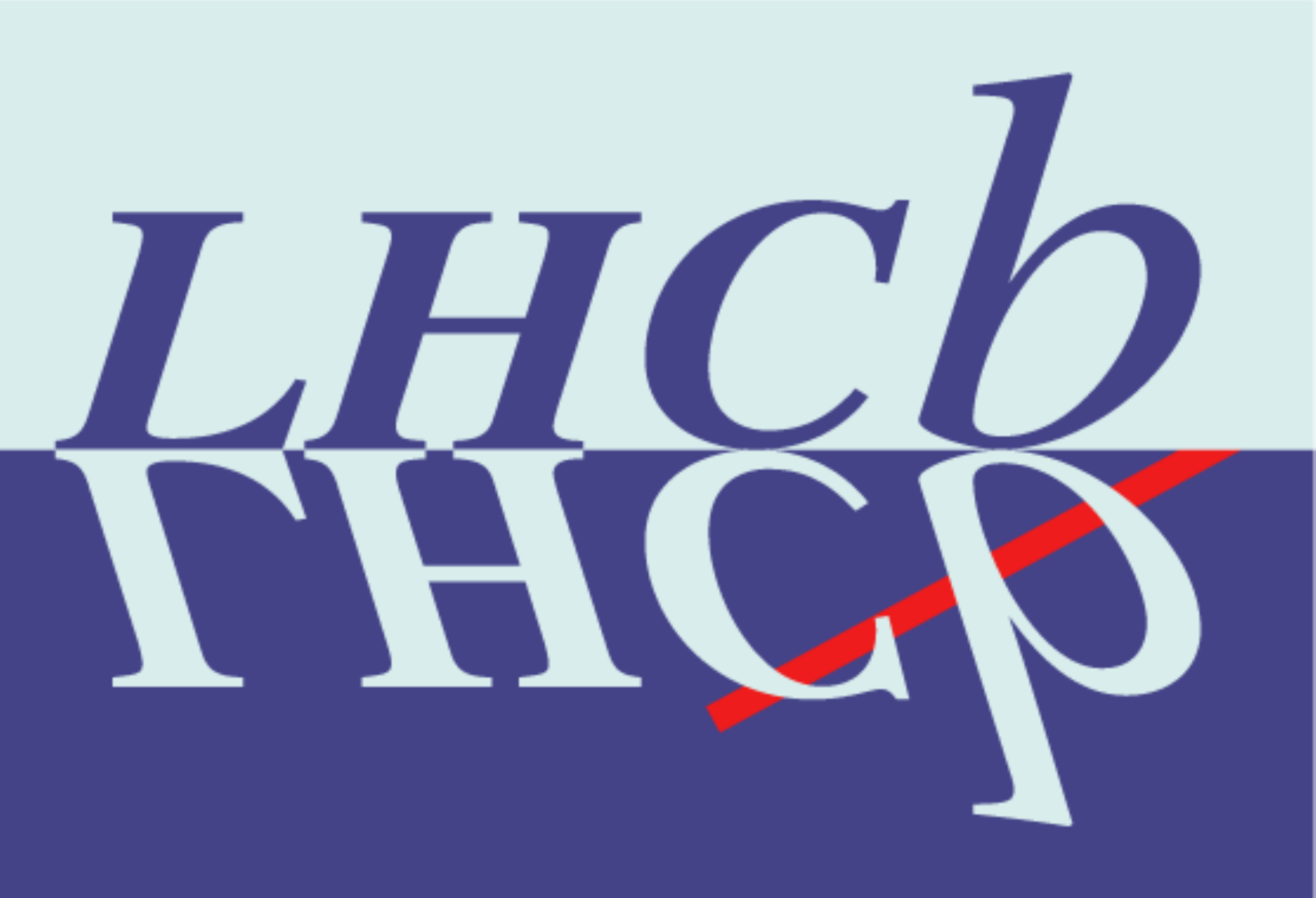}} & &}%
{\vspace*{-1.2cm}\mbox{\!\!\!\includegraphics[width=.12\textwidth]{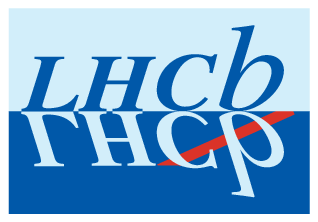}} & &}%
\\
 & & CERN-EP-2018-081 \\  
 & & LHCb-PAPER-2018-001 \\  
 & & \today \\ 
 & & \\
\end{tabular*}

\vspace*{4.0cm}

{\normalfont\bfseries\boldmath\huge
\begin{center}
  \papertitle
\end{center}
}

\vspace*{2.0cm}

\begin{center}
\paperauthors\footnote{Authors are listed at the end of this paper.}
\end{center}

\vspace{\fill}

\begin{abstract}
  \noindent
  A search for \CP and $P$ violation using triple-product asymmetries is performed with \LbTopKpipi, \LbTopKKK and \XibTopKKpiWS decays.
  The data sample corresponds to integrated luminosities of 1.0\invfb and 2.0\invfb, recorded with the \lhcb detector at centre-of-mass energies of 7\tev and 8\tev, respectively. The \CP- and $P$-violating asymmetries are measured both integrating over all phase space and in specific phase-space regions. No significant deviation from \CP or $P$ symmetry is found. The first observation of $\Lb\to\proton\Km\chi_{c0}(1P)(\to\pi^+\pi^-, K^+K^-)$ decay is also reported.

\end{abstract}

\vspace*{2.0cm}

\begin{center}
	Published in JHEP 08 (2018) 039 
\end{center}

\vspace{\fill}

{\footnotesize
\centerline{\copyright~\papercopyright. \href{\paperlicenceurl}{\paperlicence}.}}
\vspace*{2mm}

\end{titlepage}


\newpage
\setcounter{page}{2}
\mbox{~}
%
%
%
%

\cleardoublepage


\renewcommand{\thefootnote}{\arabic{footnote}}
\setcounter{footnote}{0}



\pagestyle{plain} 
\setcounter{page}{1}
\pagenumbering{arabic}


%

\section{Introduction}
\label{sec:Introduction}
The study of matter-antimatter asymmetries in \B-meson decays
contributed to establishing the validity of the Cabibbo-Kobayashi-Maskawa (CKM) mechanism for \CP violation in the Standard Model (SM).
By contrast, no \CP violation has been observed in the baryon sector to date.
However, sizeable \CP-violating asymmetries of up to 20\% are expected in certain \bquark-baryon decays~\cite{Hsiao:2014mua}, 
and a systematic study 
will either confirm the CKM mechanism in baryon decays, or will bring insights into new sources of \CP violation.
Recently the first evidence for \CP violation in
 \LbToppipipi decays has been reported by the \lhcb collaboration, with a statistical significance corresponding to 3.3 standard deviations~\cite{LHCb-PAPER-2016-030}.

 In this article, a search for \CP violation based on triple-product asymmetries in charmless \LbTopKpipi, \LbTopKKK and \XibTopKKpiWS
 decays is presented.\footnote{Unless stated otherwise, charge-conjugated modes are implicitly included throughout this article.}
 In all of these decays, the transitions are mainly
 mediated by $\bquark\to \uquark\squark\uquarkbar$ tree and $\bquark\to \squark\uquark\uquarkbar$ penguin diagrams, with a
 relative weak phase, $\arg{(V_{ub}^{}V_{us}^*/V_{tb}^{}V_{ts}^*)}$,
 that in the SM is dominated by the CKM angle $\gamma$~\cite{Gronau:2015gha}.
 With this relative phase, \CP violation could arise from the interference of these amplitudes, with the sensitivity enhanced by the rich resonant structure in \Lb and \Xibz four-body decays. The symbol \Xb is used throughout this article to refer to both \Lb and \Xibz baryons.

Asymmetries in the triple products of final-state momenta are expected to be sensitive to new physics~\cite{Golowich:1988ig,Bensalem:2002ys,Datta:2011qz}.
The triple product of final-state particle momenta in the \Xb centre-of-mass frame is defined as $\CT=\vec{p}_\proton\cdot(\vec{p}_{h_1}\times\vec{p}_{h_2})$, where $h_1 = K^-$, $h_2=\pi^+$ for the \LbTopKpipi decay, $h_1 = K^-_{\rm fast}$, $h_2 = K^+$ for the \LbTopKKK decay and $h_1 = K^-_{\rm fast}$, $h_2=\pi^+$ for the \XibTopKKpiWS decay.
The kaon labelled as ``fast (slow)'' is that with the highest (lowest) momentum among those with the same charge.
The triple product \CTbar is defined similarly for \Xbbar baryons using the momenta of the charge conjugate particles.


 Two \That-odd asymmetries are defined based on the operator \That that reverses the spin and the momentum
 of the particles~\cite{Bigi:2001sg,delAmoSanchez:2010xj,Lees:2011dx,Gronau:2011cf,Durieux:2017nps,Prasanth:2017beu}. This operator is different from the time-reversal operator, which reverses also the initial and final state. The asymmetries are defined as
\begin{align}
\label{eq:AT}
\At &= \frac{N(\CT>0)-N(\CT<0)}{N(\CT>0)+N(\CT<0)},\\
\label{eq:ATb}
\Atbar &= \frac{\kern 0.2em \overline{\kern -0.2em N}(-\CTbar>0)- \kern 0.2em \overline{\kern -0.2em N}(-\CTbar<0)}{\kern 0.2em \overline{\kern -0.2em N}(-\CTbar>0)+\kern 0.2em \overline{\kern -0.2em N}(-\CTbar<0)},
\end{align}
where $N$ and $\kern 0.2em\overline{\kern -0.2em N}$ are the numbers of \Xb and \Xbbar decays. The $P$- and \CP-violating observables are defined as
\begin{align}
\label{eq:ATV}
\aPTodd = \frac{1}{2}\left(\At +\Atbar\right),\hspace{3cm}
\aCPTodd = \frac{1}{2}\left(\At- \Atbar\right),
\end{align}
and a significant deviation from zero in these observables would indicate $P$ violation and \CP violation, respectively.
In contrast to the asymmetry between the phase-space integrated rates, \aCPTodd is sensitive to the interference of \That-even and \That-odd amplitudes and has a
different sensitivity to strong phases~\cite{PhysRevD.39.3339,Durieux:2015zwa}.
The observables \At, \Atbar, \aPTodd and \aCPTodd are, by construction, largely insensitive to \Xb/\Xbbar production asymmetries and detector-induced charge asymmetries of the final-state particles~\cite{Aaij:2014qwa}. In the present paper, these quantities are measured integrated over all the phase space and in specific phase-space regions.
%
%

\section{Detector and simulation}
\label{sec:Detector}
%
The \lhcb detector~\cite{Alves:2008zz,LHCb-DP-2014-002} is a single-arm forward
spectrometer covering the \mbox{pseudorapidity} range $2<\eta <5$,
designed for the study of particles containing \bquark or \cquark
quarks. The detector includes a high-precision tracking system
consisting of a silicon-strip vertex detector surrounding the $pp$
interaction region, a large-area silicon-strip detector located
upstream of a dipole magnet with a bending power of about
$4{\mathrm{\,Tm}}$, and three stations of silicon-strip detectors and straw
drift tubes placed downstream of the magnet. The magnetic field is reversed periodically in order to cancel detection asymmetries.
The tracking system provides a measurement of momentum, \ptot, of charged particles with
a relative uncertainty that varies from 0.5\% at 5\gevc to 1.0\% at 200\gevc.
The minimum distance of a track to a primary vertex, the impact parameter, is measured with a resolution of $(15+29/\pt)\mum$,
where \pt is the component of the momentum transverse to the beam, in\,\gevc.
Different types of charged hadrons are distinguished using information
from two ring-imaging Cherenkov detectors.
Photons, electrons and hadrons are identified by a calorimeter system consisting of
scintillating-pad and preshower detectors, an electromagnetic
calorimeter and a hadronic calorimeter. Muons are identified by a
system composed of alternating layers of iron and multiwire
proportional chambers.

The online event selection is performed by a trigger,
which consists of a hardware stage, based on information from the calorimeter and muon
systems, followed by a software stage, which applies a full event
reconstruction. Candidates are required to pass both hardware and software trigger selections.
The hardware trigger identifies the hadron daughters of the \Xb or events containing candidates generated from hard \proton\proton scattering collisions.
 The software trigger identifies four-body decays that are consistent with a $b$-hadron decay topology, and which have final-state tracks originating from a secondary vertex detached from the primary $pp$ collision point.

In the simulation, $pp$ collisions are generated using
\pythia~\cite{Sjostrand:2006za,*Sjostrand:2007gs}
 with a specific \lhcb
configuration~\cite{LHCb-PROC-2010-056}.  Decays of hadronic particles
are described by \evtgen~\cite{Lange:2001uf}, in which final-state
radiation is generated using \photos~\cite{Golonka:2005pn}. The
interaction of the generated particles with the detector, and its response,
are implemented using the \geant
toolkit~\cite{Allison:2006ve, *Agostinelli:2002hh} as described in
Ref.~\cite{LHCb-PROC-2011-006}.

\section{Candidate selection}
\label{sec:selection}
%
The analysis is based on data recorded with the \lhcb detector at centre-of-mass energies of $7\tev$ and $8\tev$, corresponding to integrated luminosities of $1.0\invfb$ and $2.0\invfb$, respectively.

The \Xb candidates are formed from combinations of tracks that originate from a good quality common vertex. 
The tracks are identified as p, \kaon or \pion candidates with loose particle identification (PID) requirements providing proton, kaon, and pion identification efficiency of 94\%, 96\% and 99\%, respectively, with a pion misidentification rate to proton (kaon) of 5\% (9\%) and a kaon misidentification rate to pion of 30\%. The proton or antiproton identifies the candidate as a \Xb baryon or \Xbbar antibaryon. Reconstructed tracks are required to have $\pt>250\mevc$ and $\ptot>1.5\gevc$, and are required to be displaced from any primary vertex.
The latter requirement is imposed by selecting tracks with $\chisqip>16$, where $\chisqip$ is the change of the primary-vertex fit \chisq when including the considered track. 
Only \Xb candidates with a transverse momentum $\pt>1.5\gevc$ are retained.
To ensure that the \Xb baryon is produced in the primary interaction, it is required that $\chisqip(\Xb)<16$, and the flight direction of the \Xb decay, calculated from its associated primary vertex, defined as that with minimum $\chisqip(\Xb)$, and the decay vertex, must align with the reconstructed particle momentum with an angle that satisfies $\cos\theta>0.9999$.

Decays of \Xb baryons to charm hadrons represent a source of background that originates from \decay{\bquark}{\cquark} transitions. Such background is vetoed by rejecting candidates with combinations of two or three final-state particles that have reconstructed
invariant masses compatible with weakly decaying charm hadron states or with the \jpsi resonance.
Among the vetoed candidates, those listed in \tablename~\ref{tab:control samples} are used for assessing systematic uncertainties and for selection criteria optimization.
\begin{table}[tb]
\centering
\small
\caption{\label{tab:control samples}Selection window for control samples used to assess systematic uncertainties and for selection criteria optimization.}
\begin{tabular}{cc}
\hline
Decay     & Selection window\\
\hline
	\decay{\Lb}{\Lc(\to p\Km\pip)\pim} & $2.23<m(p\Km\pip)<2.31\gevcc$ ($[-7.5\sigma, 4.5\sigma]$)\\
	\decay{\Lb}{\Dz(\to \Km\pip)p\pim} & $1.832<m(\Km\pip)<1.844\gevcc$ ($[-3\sigma, 3\sigma]$)\\
\hline
\end{tabular}
\end{table}
Backgrounds from a pion or a kaon misidentified as a proton originating from \Bz and \Bs decays with a \phiz or \Kstarz resonance are suppressed by vetoing the region within 10 and 70\mevcc of the \phiz and \Kstarz invariant masses,
respectively, after applying the relevant substitution of the particle mass hypotheses.

A boosted decision tree (BDT) classifier~\cite{Breiman} is used to suppress combinatorial background.
Background from other \bquark hadrons is suppressed by means of PID requirements.
The \LbTopKpipi decay, which is the final state of interest with the largest yield, is used to train the classifier, since its kinematics and topology are very similar to those of \LbTopKKK and \XibTopKKpiWS decays.
The signal training sample is obtained by subtracting the background using the \sPlot technique and a fit to the invariant mass distribution~\cite{Pivk:2004ty}.
The candidates from the sideband,
$5.85<m(\proton\Km\pip\pim)<6.40\gevcc$, are selected as the background training sample.
The discriminating variables included in the BDT are the proton transverse and longitudinal momenta \pt and $p_z$;
the impact parameter of the \kaon and \pion candidate tracks with respect to the \Xb primary vertex;
the \chisq of the \Xb decay vertex fit; the angle between the \Xb momentum and its flight direction;
the \Xb $\chisqip$;
the asymmetry between the transverse momentum of the \Xb and that of the charged tracks contained in a region defined as $\sqrt{\Delta\eta^2+\Delta\phi^2}<1.0$, where $\Delta\eta$ ($\Delta\phi$) is the difference of pseudorapidity (azimuthal angle) between the candidate and the charged tracks. 
The most important discriminating variables are the proton transverse and longitudinal momentum, and the angle between the \Xb momentum and its flight direction. No correlation is found between the discriminating variables or between the BDT output and the reconstructed \bquark-baryon candidate mass.
The signal and background training samples are divided into three statistically independent subsamples with equal number of candidates, on which $k$-fold cross-validation is applied~\cite{arlot2010}.
The BDT selection criteria are optimised by maximising $S/\sqrt{S + B}$,
where $S$ ($B$) is the expected signal (background) yield.
The expected yield is estimated using $S=\epsilon_{S}S_0$ ($B=\epsilon_{B}B_0$), where the signal (background) efficiency $\epsilon_{S}$ ($\epsilon_{B}$) of each BDT selection requirement is evaluated using \LbTopKpipi (data sideband) control samples; the reference signal (background) yield, $S_0$ ($B_0$), is obtained from a fit to the reconstructed invariant mass in the range $[5.5-5.7]\gevcc$ before applying the BDT selection.
%

The  \decay{\Lb}{\proton\Dz(\to\Km\pip)\pim} sample is employed to optimise the PID selection since the momentum and pseudorapidity distributions of its final-state particles are
similar to those of \LbTopKpipi, \LbTopKKK and \XibTopKKpiWS decays.
The figure of merit that is maximised is defined as
\begin{align}
\label{eq:FoM}
\mathcal{S}_{\textrm{PID}}= \frac{\varepsilon_S(\text{PID}) \cdot N_S}{\sqrt{\varepsilon_S(\text{PID}) \cdot N_S + \varepsilon_B(\text{PID}) \cdot N_B}},
\end{align}
where the signal and background efficiencies of the PID selection criteria,
$\varepsilon_S(\text{PID})$ and $\varepsilon_B(\text{PID})$, respectively, are determined using the \decay{\Lb}{\proton\Dz(\decay{}{\Km\pip})\pim} sample;
$N_S$ ($N_B$) is the number of signal (background) candidates after applying the BDT selection. 
Multiple candidates are reconstructed in less than 1\% of the selected events, and in such cases a single candidate is retained with a random but reproducible choice.

There are three main categories of background considered in the optimization process. Background from partially reconstructed decays is localised in the region at low invariant mass, and originates from \decay{\Lb}{p\pi^+K^-\rho^-(\decay{}{\pim\piz})}, \decay{\Lb}{ p\pi^+\pi^-\Kstarm(\decay{}{\Km\piz})} and similar decays, where the \piz meson is not reconstructed. The background from misidentified final-state particles, called cross-feed in the following, consists of four-body \Lb, \Bz and \Bs decays, where one of them is reconstructed with the wrong mass hypothesis. The combinatorial background results from random combinations of tracks in the event.

%
\section{Measurement of the $\boldsymbol \CP$-violating asymmetries}
\label{sec:analysis-cpv}
For each signal mode, the selected data sample is split into four subsamples according to the \Xb or \Xbbar flavour and the sign of \CT or \CTbar.
Simulated events and the \decay{\Lb}{\Lc(p\Km\pip)\pim} control sample indicate that the reconstruction efficiencies for candidates with $\CT>0$ ($-\CTbar>0$) and $\CT<0$ ($-\CTbar<0$) are equal, within statistical uncertainties.
For each final state, a simultaneous maximum likelihood fit to the $m(p\Km h^+h^-)$ distribution of the four subsamples is used to determine the number of signal and background yields and the asymmetries \At and \Atbar.
The $P$- and \CP-violating asymmetries, \aPTodd and \aCPTodd, are then obtained according to Eq.~(\ref{eq:ATV}).

\begin{figure}[tb]
\centering
\small
\includegraphics[width=0.45\textwidth]{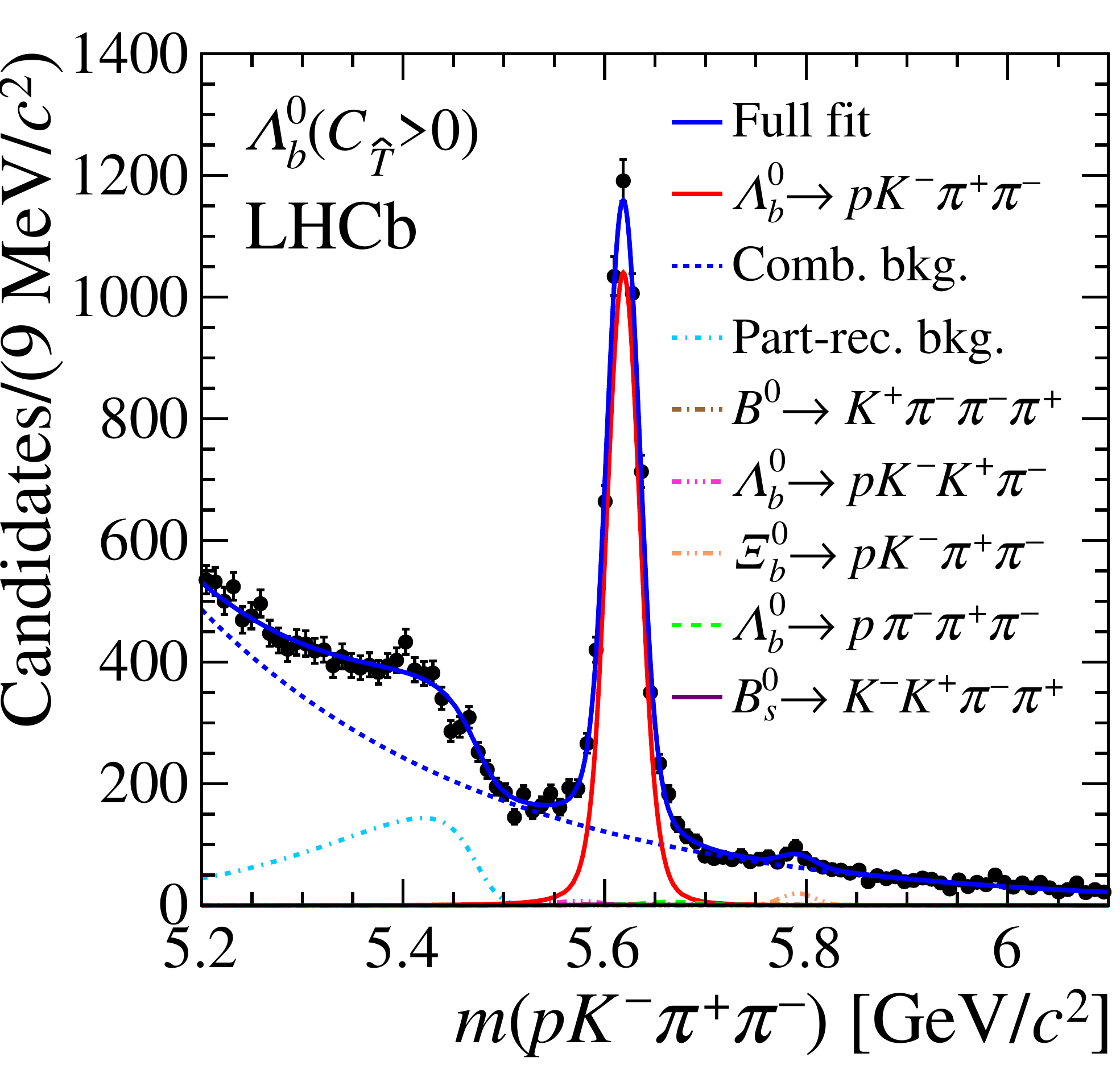}
\includegraphics[width=0.45\textwidth]{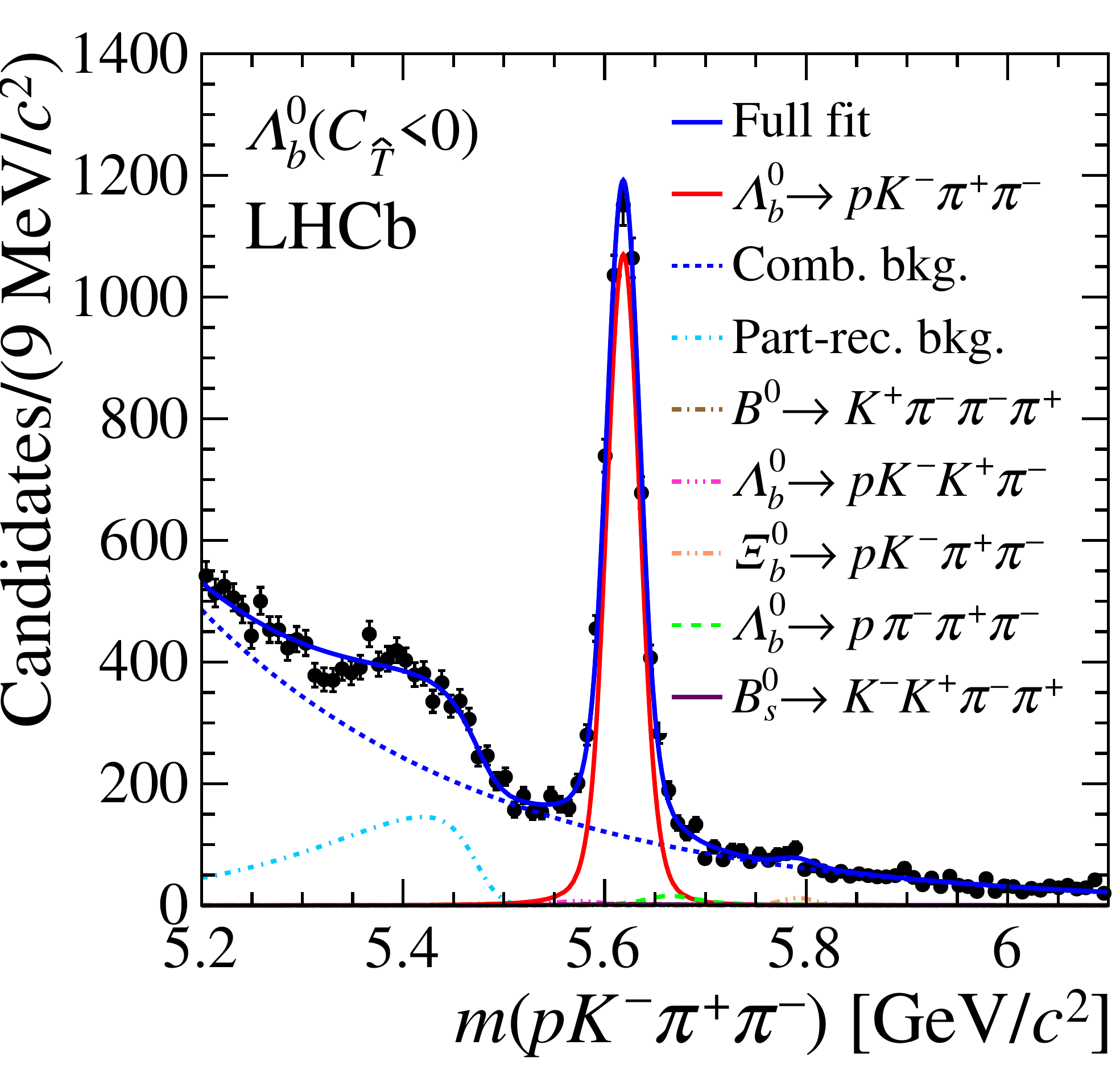}
\includegraphics[width=0.45\textwidth]{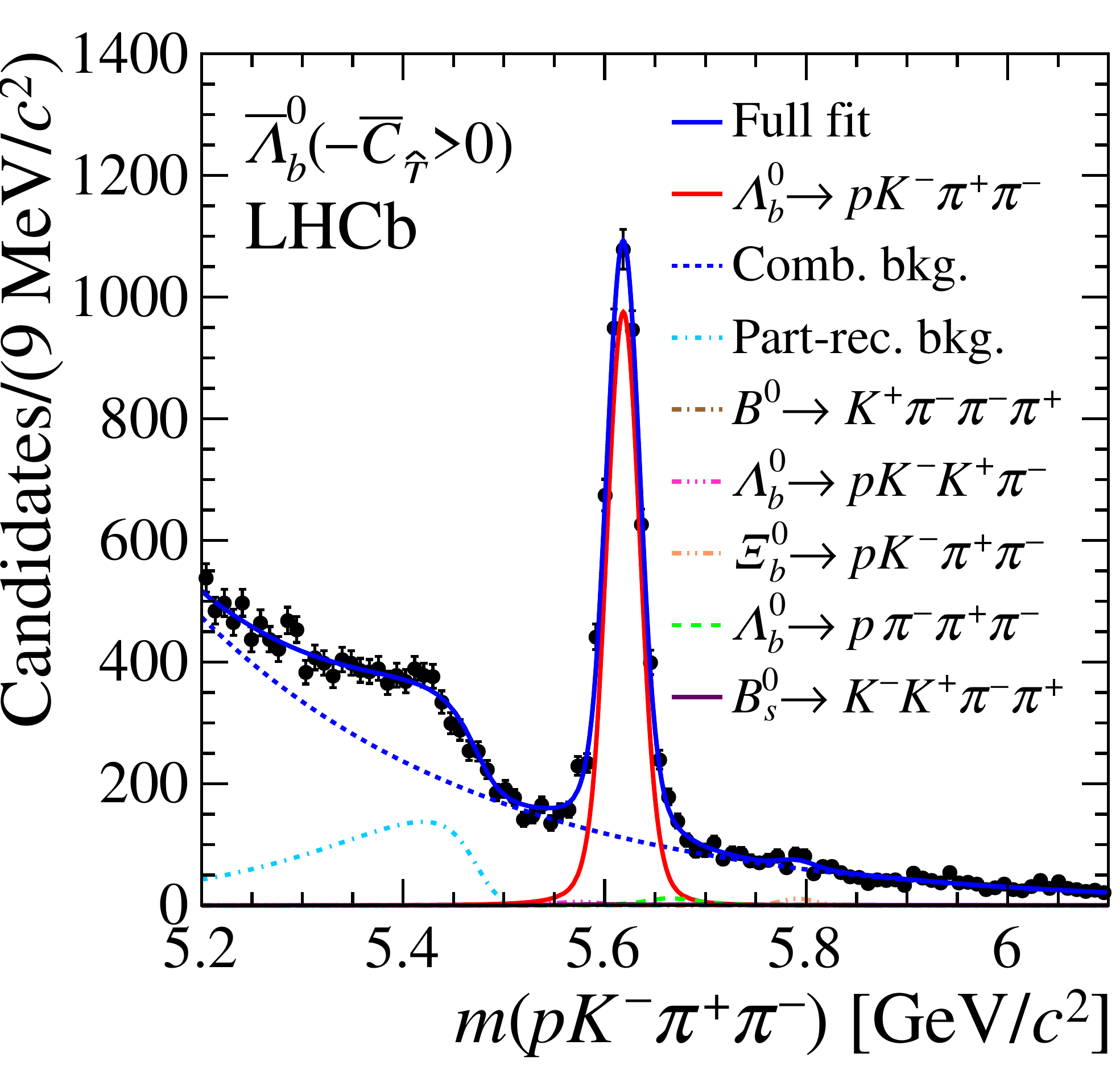}
\includegraphics[width=0.45\textwidth]{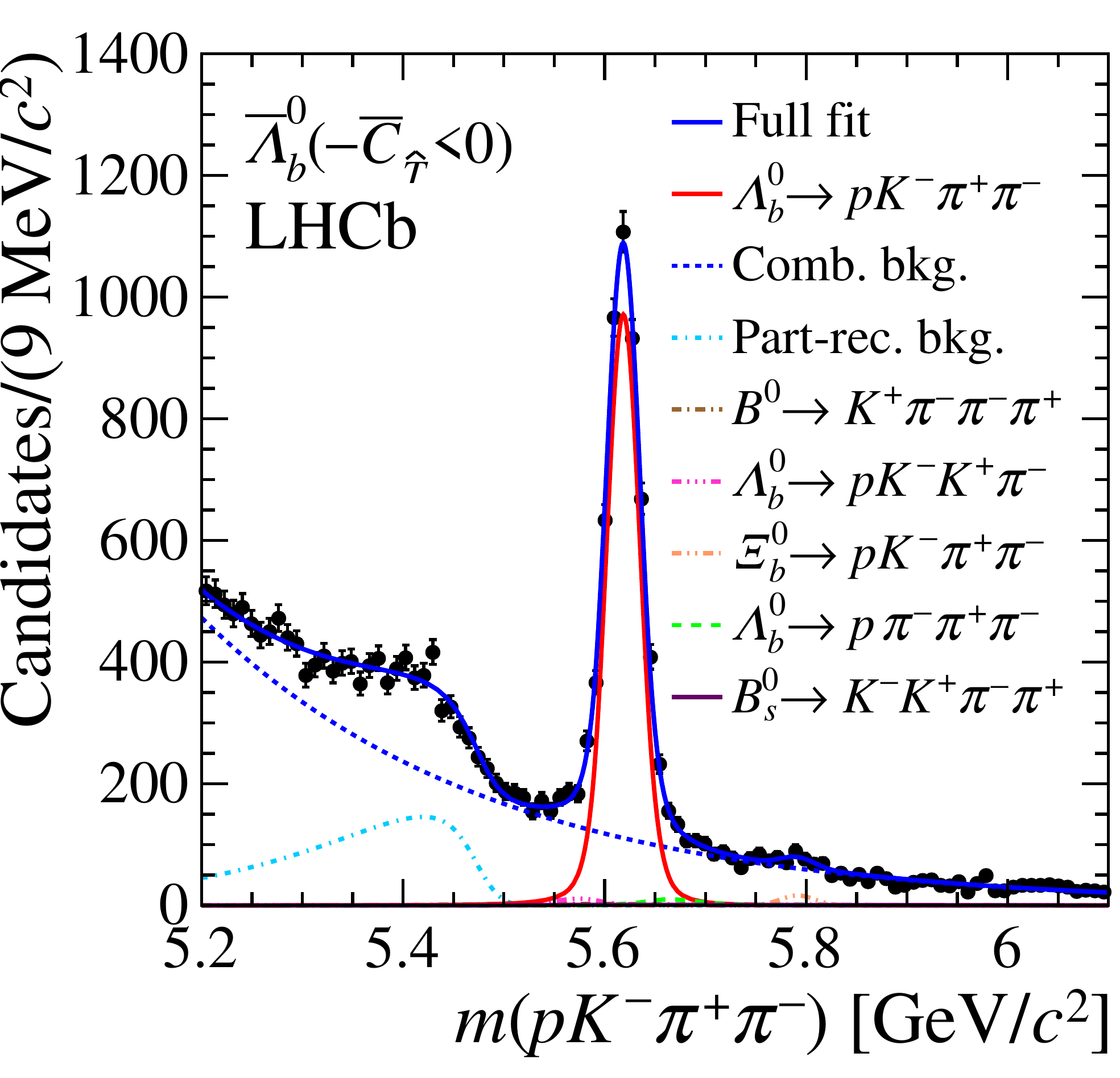}
\caption{\small\label{fig:Lb2pK2pi_asym}Distributions of the $p\Km\pip\pim$ invariant mass in the four samples defined by the \Lb (\Lbbar) flavour and the sign of \CT (\CTbar). The results of the fit are overlaid as described in the legend.
The contribution of the cross-feeds to the fit results is barely visible but is found to be nonnegligible.} 
\end{figure}

\begin{figure}[tb]
\centering
\small
\includegraphics[width=0.45\textwidth]{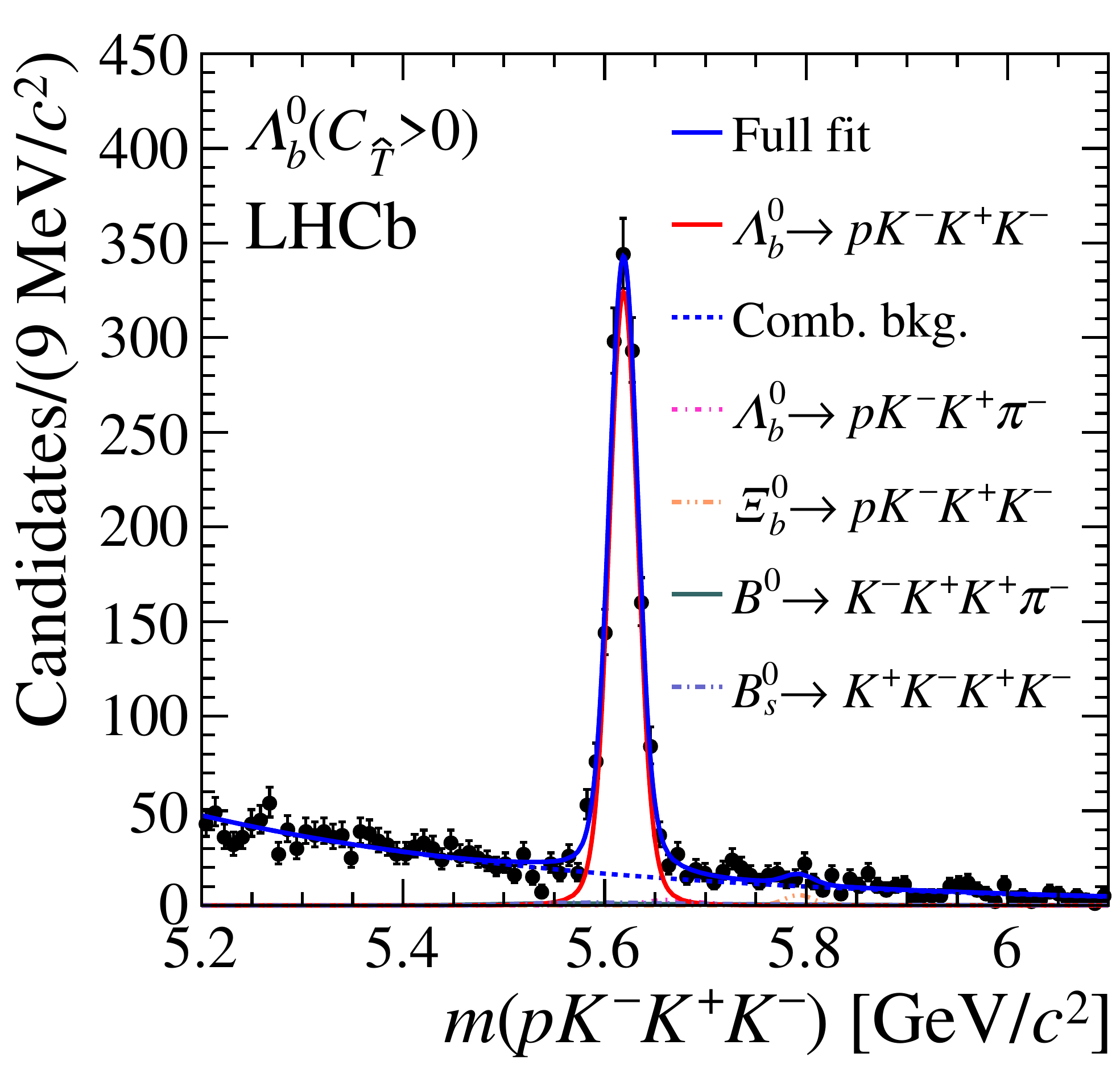}
\includegraphics[width=0.45\textwidth]{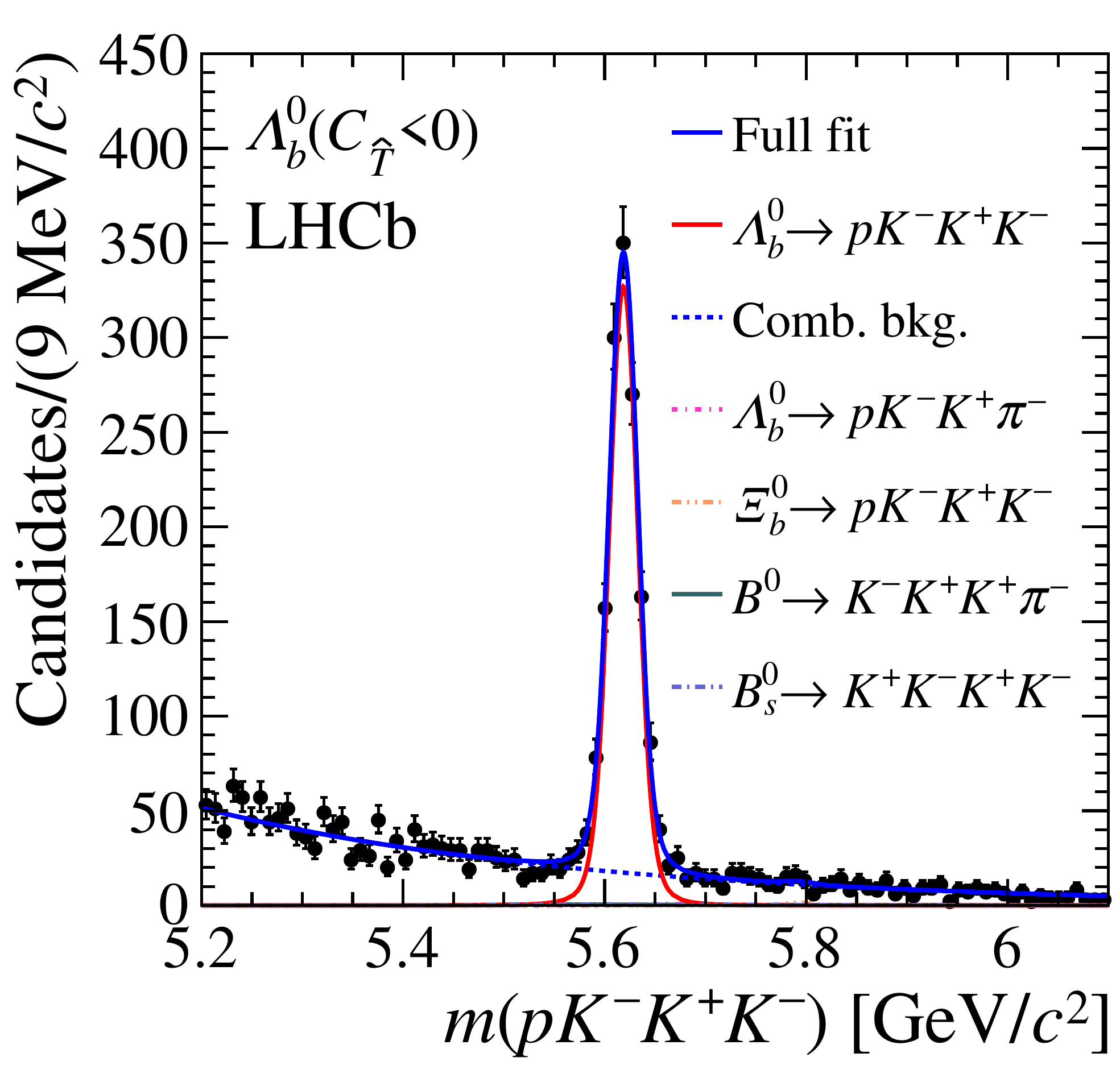}
\includegraphics[width=0.45\textwidth]{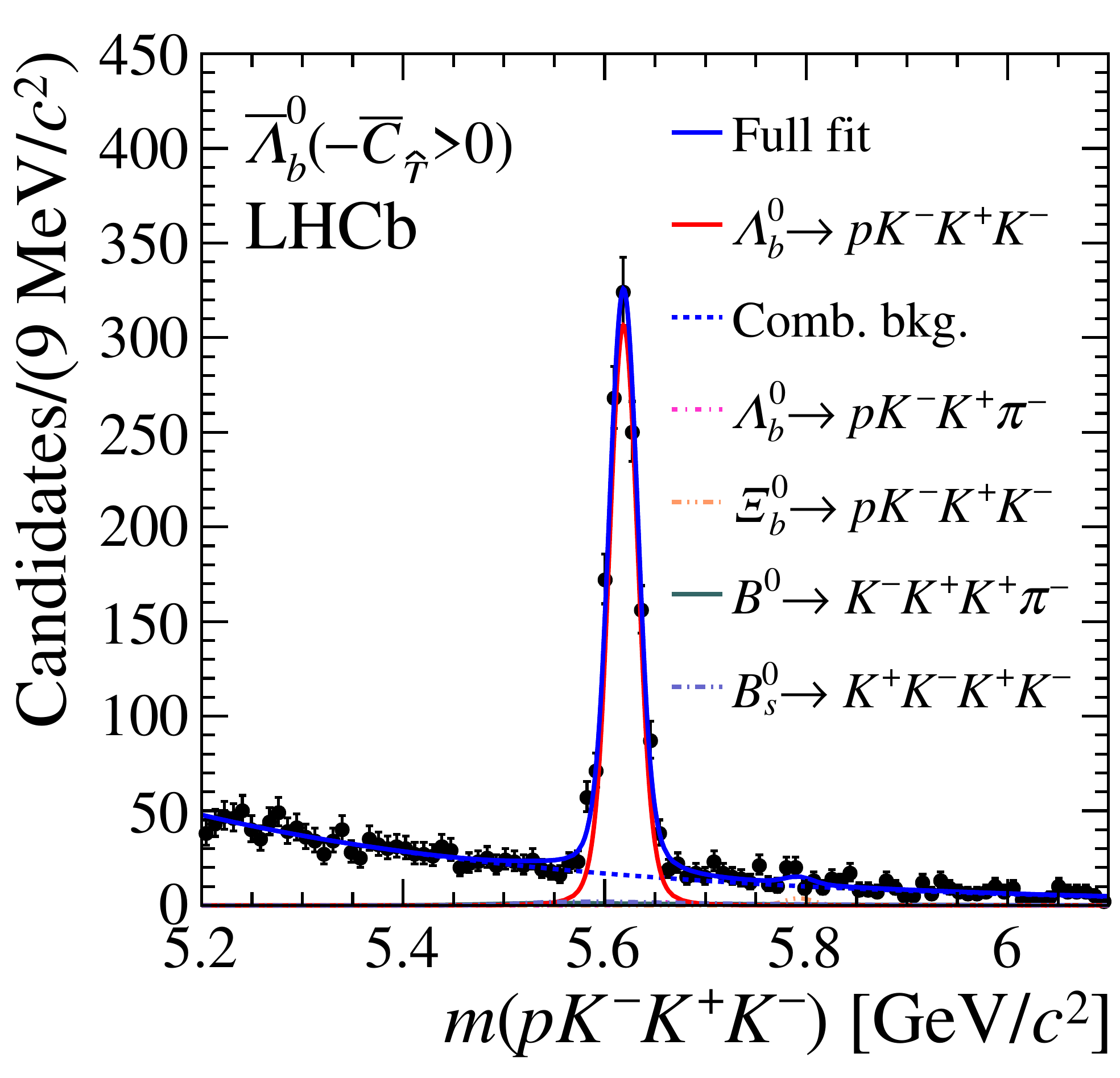}
\includegraphics[width=0.45\textwidth]{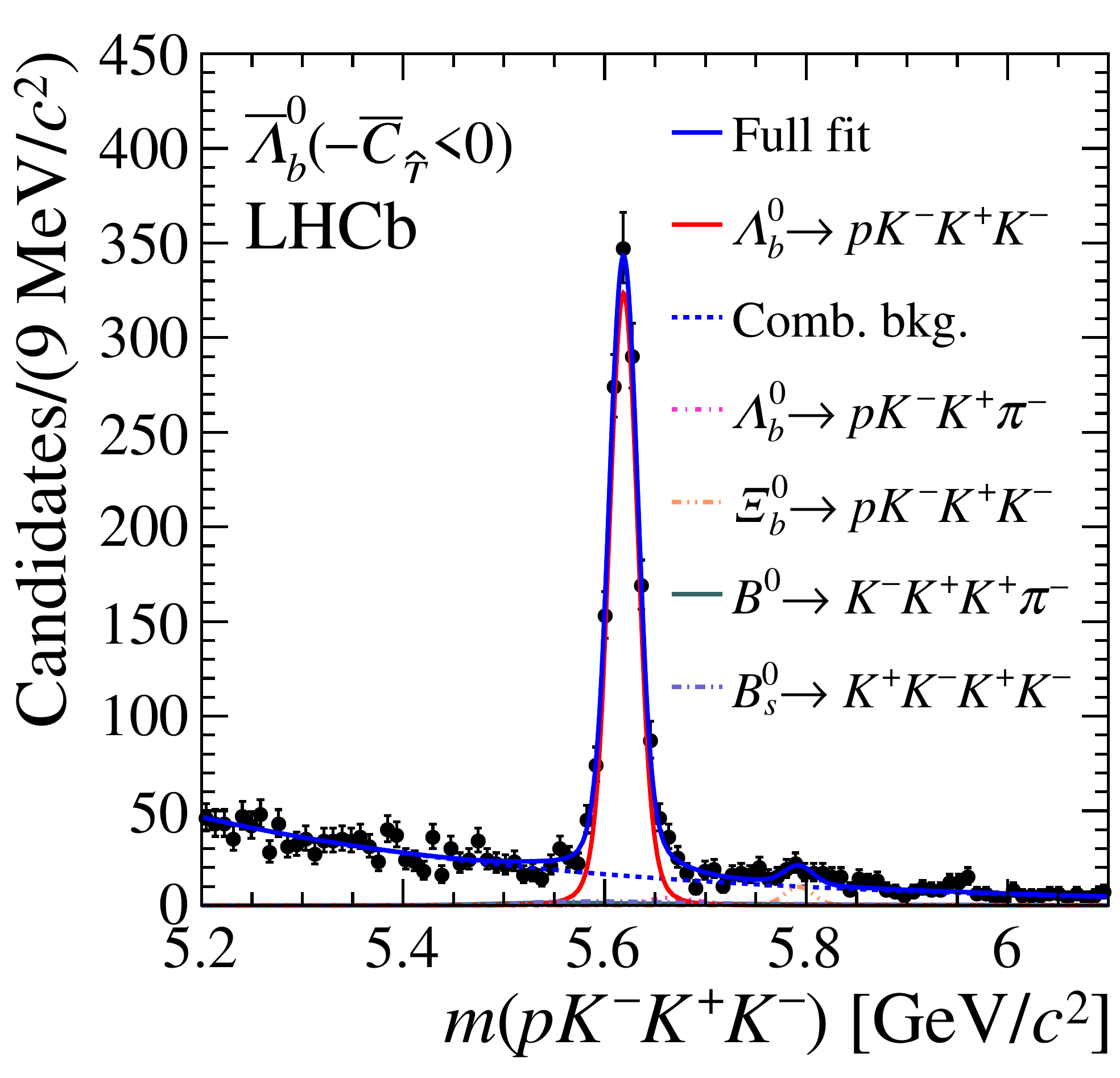}
\caption{\small\label{fig:Lb2p3K_asym}Distributions of the $p\Km\Kp\Km$ invariant mass in the four samples defined by the \Lb (\Lbbar) flavour and the sign of \CT (\CTbar). The results of the fit are overlaid as described in the legend.
The contribution of the cross-feeds to the fit results is barely just visible but is found to be nonnegligible.}
\end{figure}

\begin{figure}[tb]
\centering
\small
\includegraphics[width=0.45\textwidth]{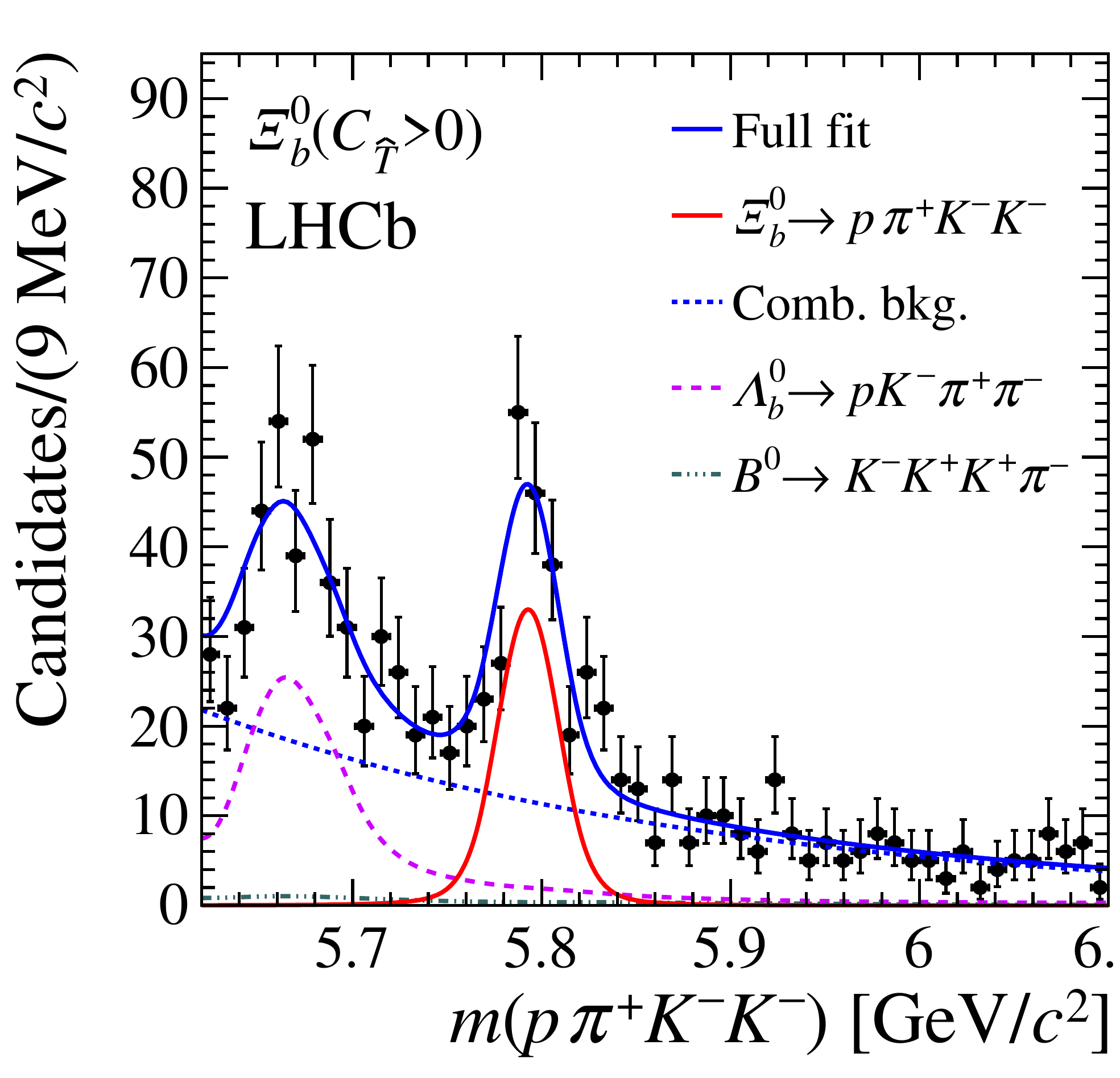}
\includegraphics[width=0.45\textwidth]{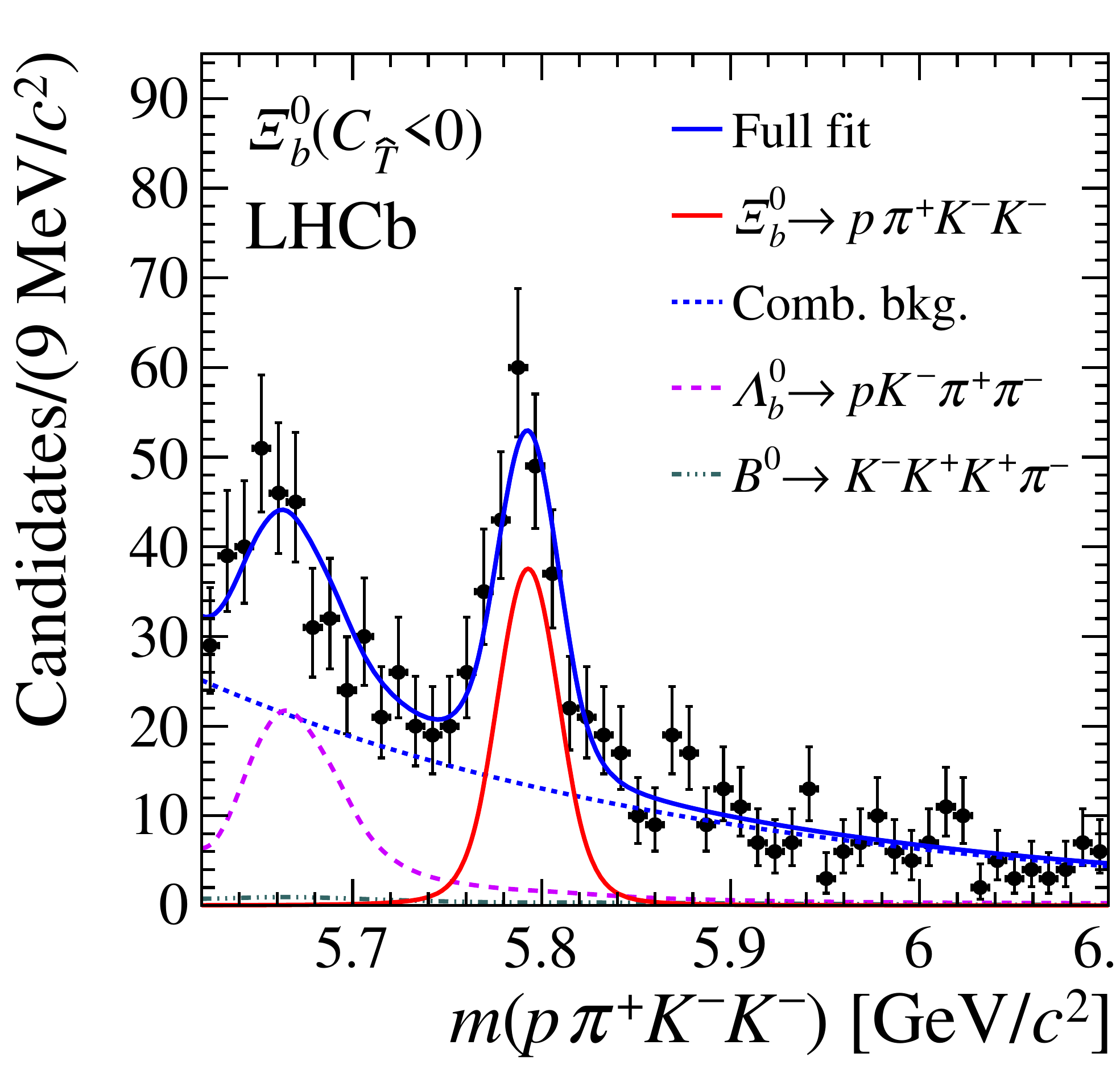}
\includegraphics[width=0.45\textwidth]{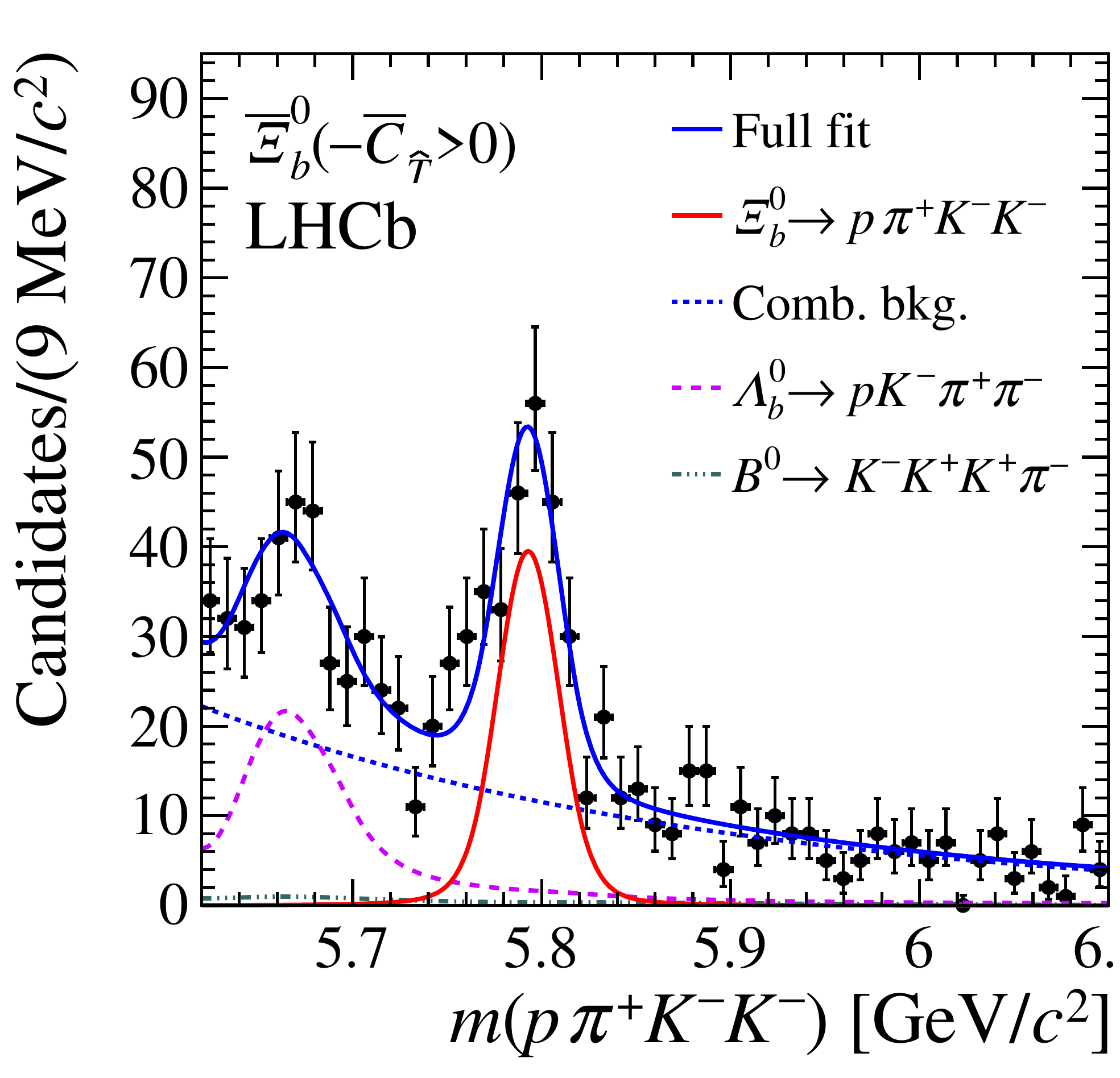}
\includegraphics[width=0.45\textwidth]{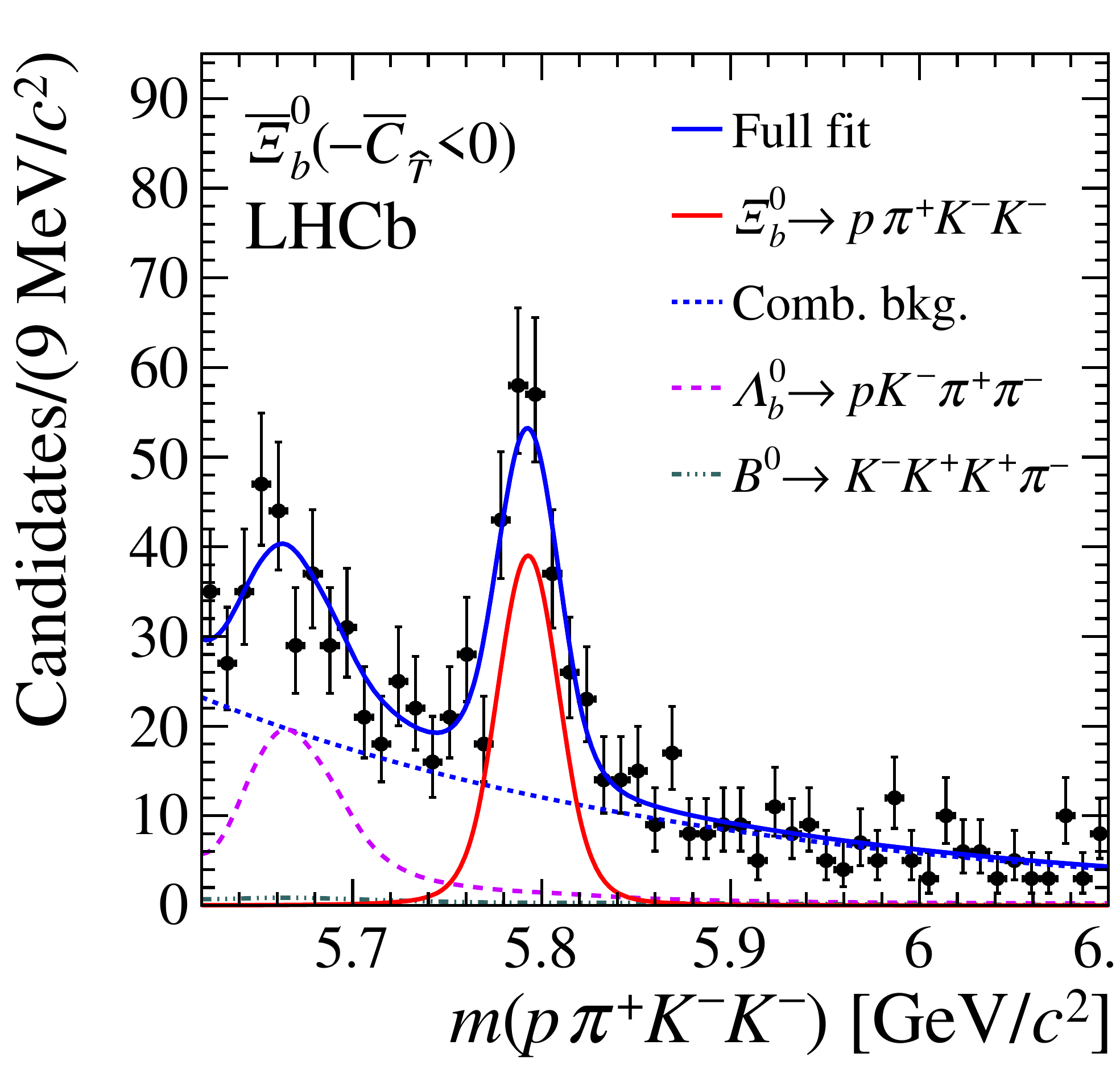}
\caption{\small\label{fig:Xib2p2KpiWS_asym}Distributions of the $p\Km\Km\pip$ invariant mass in the four samples defined by the \Xibz (\Xibbarz) flavour and the sign of \CT (\CTbar). The results of the fit are overlaid as described in the legend.
The contribution of the \decay{\Bz}{\Km\Kp\Kp\pim} cross-feed to the fit results is barely visible but is found to be nonnegligible.}
\end{figure}

The invariant-mass distribution of the \Xb signal is modelled by the sum of two Crystal Ball
functions~\cite{Skwarnicki:1986xj} that share the peak value and width but have tails on opposite sides of the peak. The parameters related to the tails and the relative fraction of the two Crystal Ball functions are determined from fits to simulated samples, and are fixed in fits made to data. The \Xibz signal is also visible in the $m(p\Km\pip\pim)$ and $m(p\Km\Kp\Km)$ invariant-mass distributions,
and its peak value is fitted by imposing a Gaussian constraint using the known value of the mass difference of the \Xibz and \Lb baryons, $174.8\pm2.5 \mevcc$~\cite{PDG2014}.
The combinatorial background distribution is modelled by an exponential function with the rate parameter determined from the data.
Partially reconstructed \Lb decays are described by a threshold function~\cite{Albrecht:1990am} convolved with a Gaussian function to account for resolution effects, the parameters of which are determined from the fit.
The shapes of cross-feed backgrounds are modelled using non-parametric functions~\cite{Cranmer:2000du} based on simulated events.
The fit results for \LbTopKpipi, \LbTopKKK, and \XibTopKKpiWS decays are shown in Figs.~\ref{fig:Lb2pK2pi_asym},~\ref{fig:Lb2p3K_asym}, and~\ref{fig:Xib2p2KpiWS_asym}, respectively.
The signal yields, $19877\pm195$, $5297\pm83$, and $709\pm45$, respectively, are compatible with the previously measured branching fractions~\cite{Aaij:2017pgy}, once the selection efficiencies are taken into account.
%
In the \LbTopKpipi and the \LbTopKKK decay modes, signals consistent with the $\chi_{c0}(1P)$ charmonium resonance are observed in the \pip\pim and the $\Kp K_{\rm fast}^-$ invariant-mass distributions, which are shown in Fig.~\ref{fig:xic02} in Appendix~\ref{app:xic02}. 
The signal yield 
and the corresponding statistical uncertainty 
for the $\Lb\to\proton\Km\chi_{c0}(1P)(\to\pi^+\pi^-)$ decay is $336\pm25$, and for the
$\Lb\to\proton\Km\chi_{c0}(1P)(\to K^+K^-)$ decay is $332\pm23$, representing the first observation of these decays.
The $\decay{\Lb}{\proton\Km}\chi_{c0}(1P)$ candidates have an identical final state to the \LbTopKpipi and \LbTopKKK signal decays and can potentially contribute to \CP violation. These candidates are retained, together with the charmless 4-body decays, for the measurements of the asymmetries described below. Similar decays from \decay{\Lb}{\proton\Km\jpsi} with \decay{\jpsi}{\pip\pim} are removed due to the significant background from misidentified \decay{\jpsi}{\mumu} decays.

Two different approaches have been used to search for $P$ and \CP violation: a measurement integrated over the phase space and measurements in specific phase-space regions.
The results of the first approach are obtained by fitting the full data sample and found to be compatible with $P$ and \CP symmetries, as shown in Table~\ref{tab:PHSP_Asym}.

%
%

%
%

\begin{figure}[tb]
\centering
\small
\includegraphics[width=0.45\textwidth]{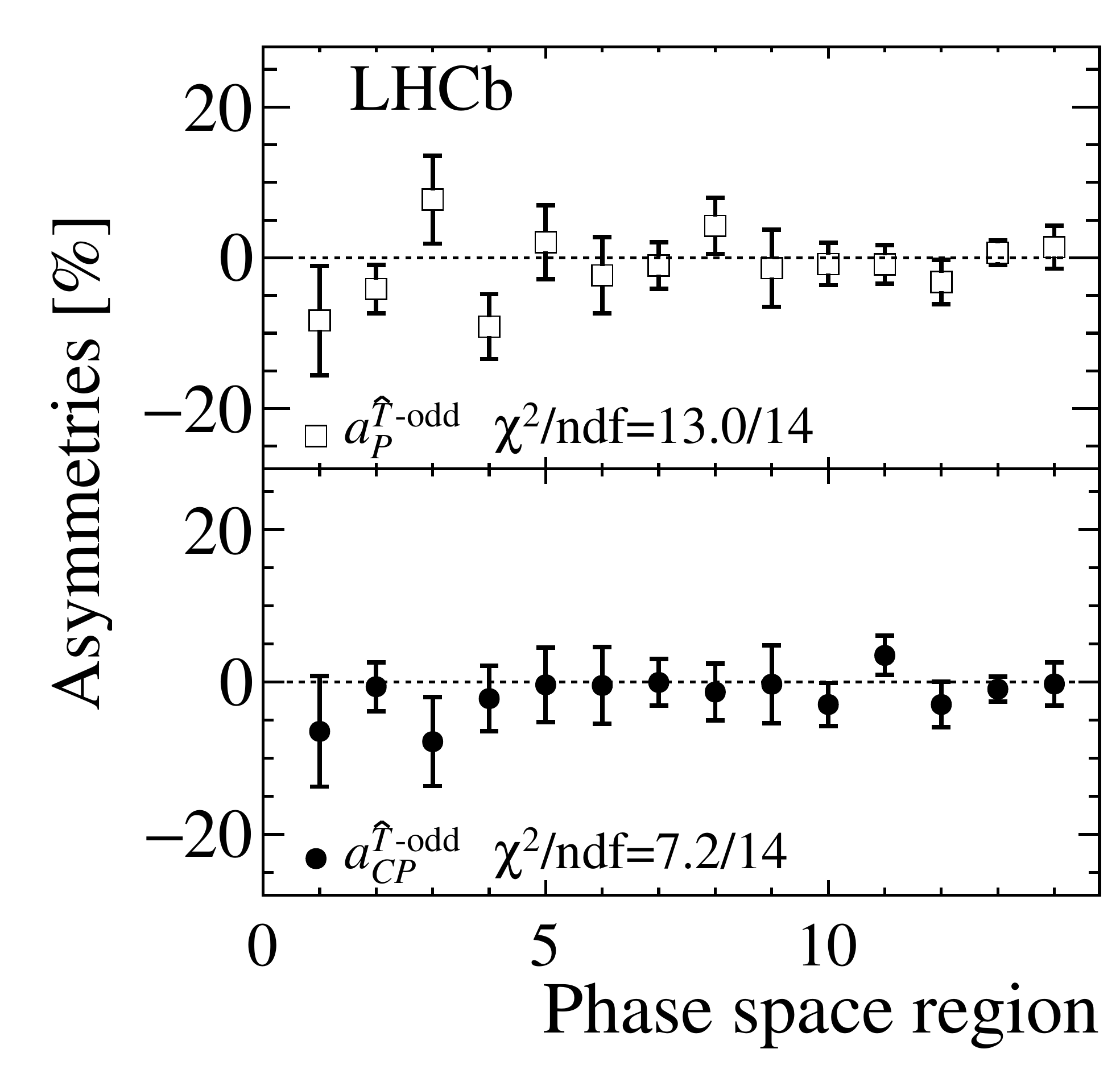}
\includegraphics[width=0.45\textwidth]{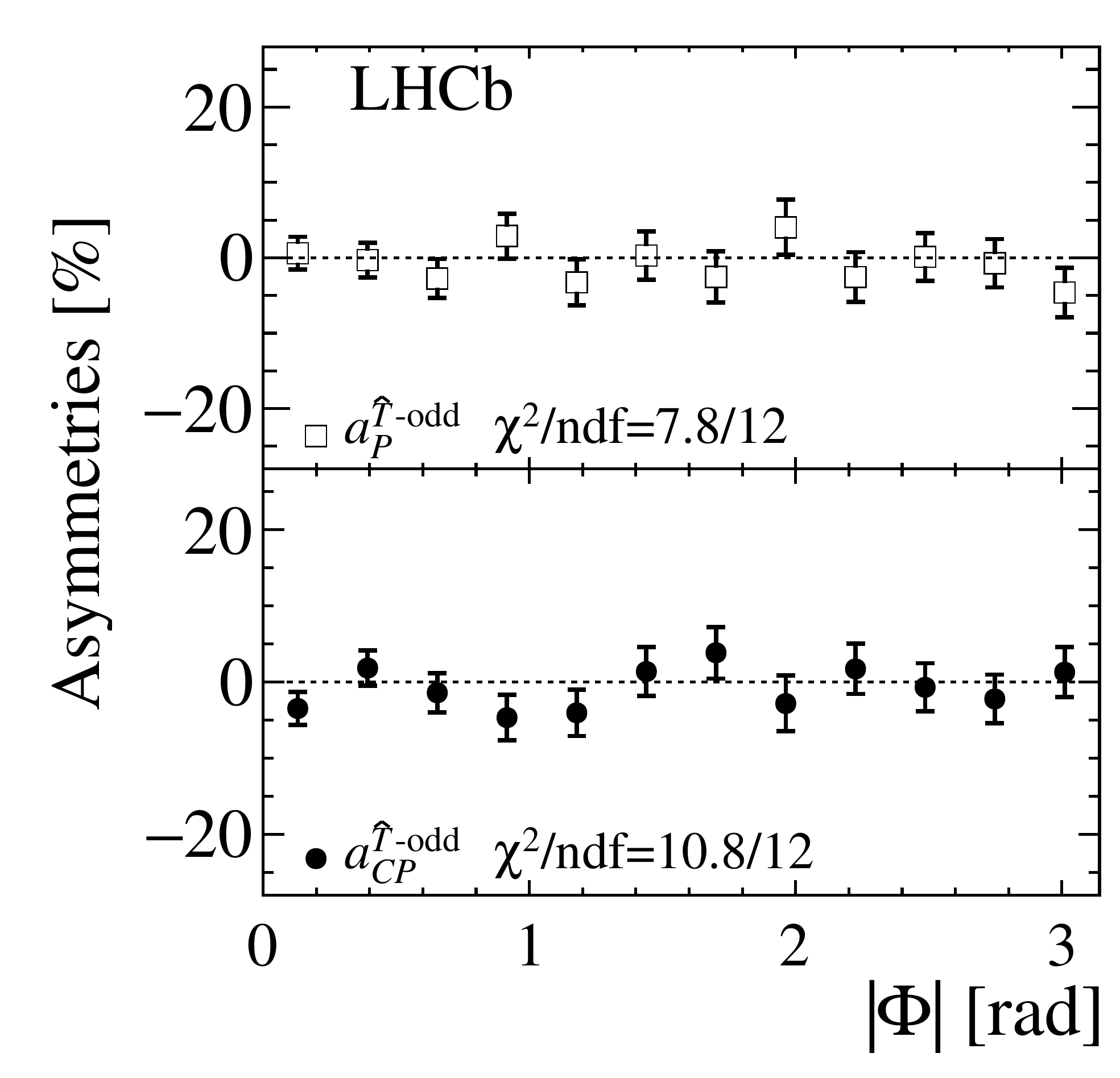}
\caption{\small\label{fig:Lb2pK2pi_asym_phsp}The asymmetries using binning schemes (left) A and (right) B for the \LbTopKpipi decay.
For \aPTodd (\aCPTodd), the values of the \chisq/ndf for the $P$-symmetry (\CP-symmetry)  hypothesis, represented by a dashed line, are quoted.}
\end{figure}
\begin{figure}[tb]
\centering
\small
\includegraphics[width=0.45\textwidth]{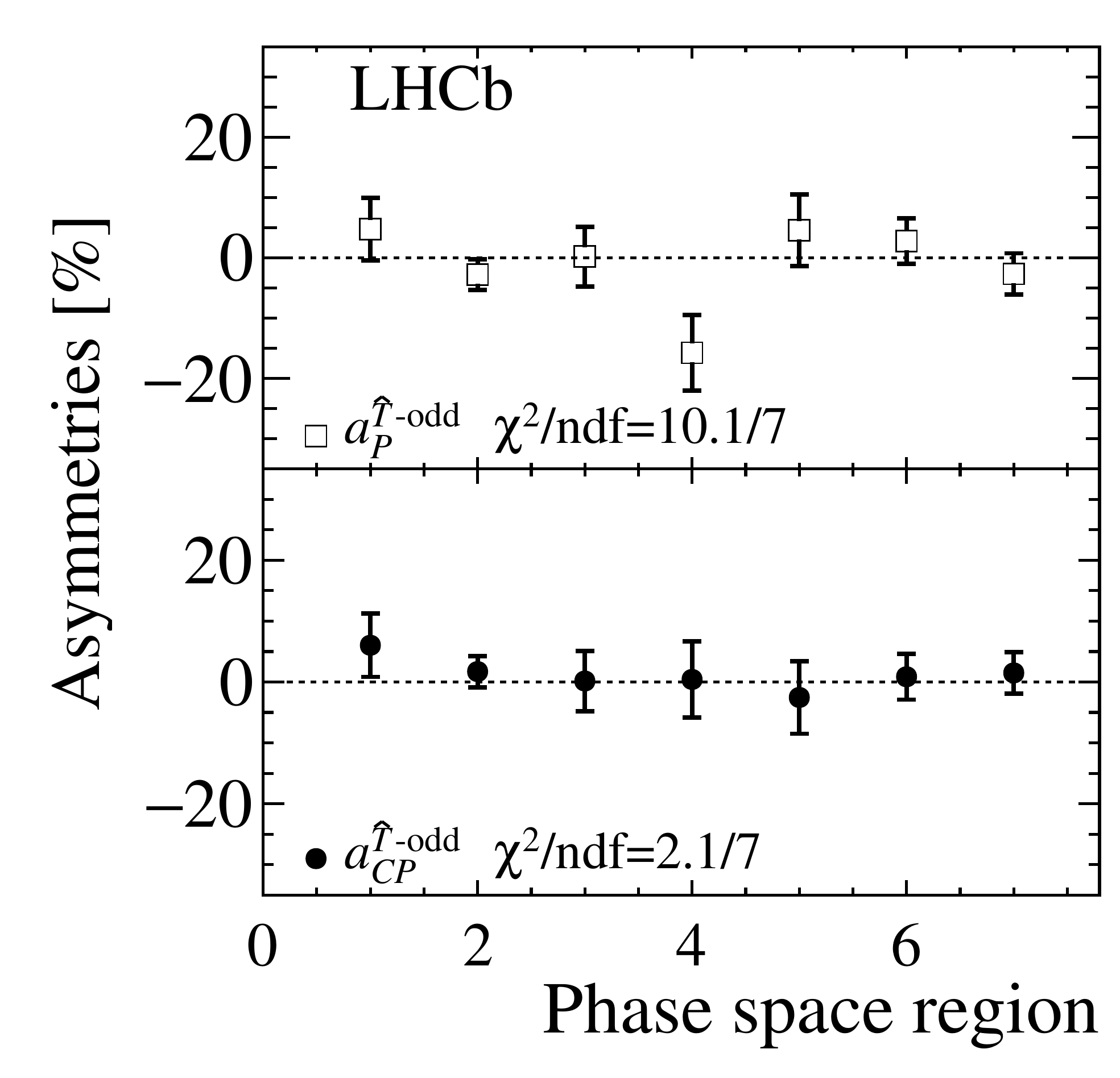}
\includegraphics[width=0.45\textwidth]{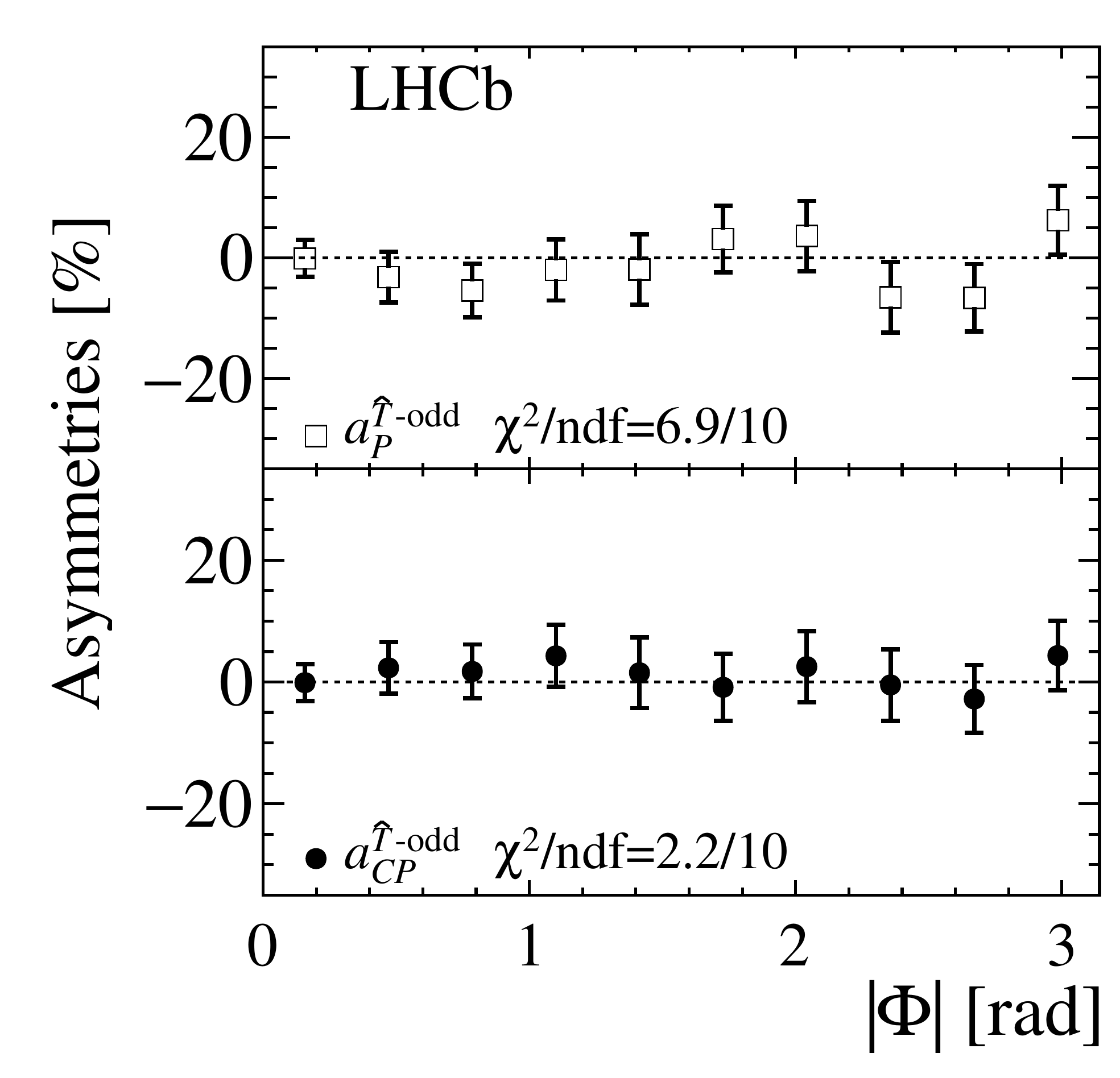}
\caption{\small\label{fig:Lb2p3K_asym_phsp}The asymmetries using binning schemes (left) C and (right) D for \LbTopKKK decay.
For \aPTodd (\aCPTodd), the values of the \chisq/ndf for the $P$-symmetry (\CP-symmetry) hypothesis, represented by a dashed line, are quoted.}
\end{figure}

The \CP-violating asymmetries may vary over the phase space due to the interference between resonant contributions.
Therefore, measurements in specific phase-space regions may have better sensitivity to \CP violation.
In order to avoid biases, the binning schemes used to divide up the phase space were chosen before examining the data.
Two binning schemes are used for the \LbTopKpipi (\LbTopKKK) decay. Schemes A and B (C and D) are designed to isolate regions of phase space according to the dominant resonant contributions and to exploit the potential interference of contributions as a function of the angle $\Phi$ between the decay planes formed by the $\proton\Km$ ($\proton K_{\rm fast}^-$) and the $\pip\pim$ ($\Kp K_{\rm slow}^-$) systems, respectively.
Scheme A (C) is defined in Table~\ref{tab:PHSP_Def_SchemeA} (\ref{tab:PHSP_Def_SchemeC}) in Appendix~\ref{app:phase_space}, while scheme B (D) has twelve (ten) nonoverlapping bins of width $\pi/12$ ($\pi/10$) in $|\Phi|$. The size of the bins, and the resulting statistical uncertainty, is chosen to have sensitivity at the level of a few percent. The same fit model used for the integrated measurement is employed to fit each phase-space region.
The distribution of asymmetries for the \LbTopKpipi (\LbTopKKK) decay is shown in Fig.~\ref{fig:Lb2pK2pi_asym_phsp} (\ref{fig:Lb2p3K_asym_phsp}), and the results are reported in Table~\ref{tab:PHSP_Asym_SchemeAB} (\ref{tab:PHSP_Asym_SchemeCD}) in Appendix~\ref{app:phase_space}.

The compatibility with the \CP-symmetry ($P$-symmetry) hypothesis is tested for each scheme individually by means
of a \chisq test,
where the \chisq is defined as $R^TV^{-1}R$, with $R$ the array of \aCPTodd (\aPTodd) measurements and $V^{-1}$ the inverse of the covariance matrix, which is the sum of the statistical and systematic covariance matrices.
An average systematic uncertainty, discussed in Section~\ref{sec:syst}, is assumed for all bins. The statistical uncertainties are considered uncorrelated among the bins, while systematic uncertainties are assumed to be fully correlated.
The results are consistent with the \CP-symmetry hypothesis with a $p$-value of $0.93$ (0.55), based on \chisq/ndf$=7.2/14$ ($10.8/12$) for scheme A (B) and a $p$-value of $0.95$ (0.99), based on \chisq/ndf$=2.1/7$ ($2.2/10$) for scheme C (D). A similar \chisq test is performed on the \aPTodd measurements. The results are consistent with the $P$-symmetry hypothesis with a $p$-value of $0.53$ (0.80), based on \chisq/ndf$=13.0/14$ ($7.8/12$) for scheme A (B) and a $p$-value of $0.18$ (0.73), based on \chisq/ndf$=10.1/7$ ($6.9/10$) for scheme C (D).
%
\begin{table}[ht]
\centering
\small
\caption{\label{tab:PHSP_Asym} Measurements of the \CP- and $P$-violating observables \aCPTodd and \aPTodd, together with their statistical and systematic uncertainties.
}
\begin{tabular}{crrr}
\hline
	& \LbTopKpipi	& \LbTopKKK	& \XibTopKKpiWS \\
\hline
\aPTodd (\%)	&	$-0.60\pm0.84\pm0.31$	&	$-1.56\pm1.51\pm0.32$ & $-3.04\pm5.19\pm0.36$ \\
\aCPTodd (\%)	&	$-0.81\pm0.84\pm0.31$	&	$1.12\pm1.51\pm0.32$	&	$-3.58\pm5.19\pm0.36$ \\
\hline
\end{tabular}

\end{table}

%
%

\section{Evaluation of systematic uncertainties}
\label{sec:syst}
The sources of systematic uncertainty and their relative contributions to the total uncertainty are listed in Table~\ref{tab:Systematics}.
The main source of systematic uncertainty is due to the experimental reconstruction and analysis technique, which could introduce potential biases in the measured asymmetries.
This is tested by measuring the asymmetry $\aCPTodd(\Lc\pim)$ for the Cabibbo-favoured $\Lb\to\Lc\pim$ decay mode, where negligible \CP violation is expected.
The measured asymmetry is consistent with zero with a statistical uncertainty of $0.31\%$, which is assigned as a systematic uncertainty for \aCPTodd for the integrated measurement over the full phase space. The systematic uncertainty on \aPTodd is identical to that on \aCPTodd, as follows from Eq.~\ref{eq:ATV}.

To assess the systematic uncertainty for the measurements in regions of the phase space, the $\Lb\to\Lc(\to p\Km\pip)\pim$ control sample is split in ten bins of the angle $\Phi$ between the decay planes of $p\Km$ and $\pip\pim$.
The resulting distribution of \aCPTodd is fitted with various models, all of which give results consistent with no asymmetry with a statistical precision of $0.6\%$. This statistical precision is assigned as a systematic uncertainty in each bin of the different binning schemes A, B, C and D.

The reconstruction efficiencies for signal candidates of opposite sign of \CT are identical within statistical uncertainties of the control sample and the signal MC, and likewise for \CTbar, which indicates that the detector and the reconstruction technique do not bias the asymmetry measurements. Similarly, the reconstruction efficiencies over $|\Phi|$ and four-body phase space are also identical for events with opposite sign of \CT and \CTbar. For the measurements of the triple products \CT and \CTbar, the systematic uncertainty from detector-resolution effects, which could introduce a migration of signal decays between the bins, is estimated from simulated samples of \LbTopKpipi, \LbTopKKK and \XibTopKKpiWS decays, where neither $P$- nor \CP-violating effects are present. The difference between the reconstructed and generated asymmetry is taken as systematic uncertainty and is less than $0.05\%$ in all cases.

The systematic uncertainties related to the choice of model for the signal and background components of the fits are evaluated by using alternative models that have comparable fit quality.
The signal shape is varied by weighting the simulated sample with the PID efficiencies determined from data in order to account for possible discrepancies between data and simulation.
The power and the threshold parameters of the empirical function for the partially reconstructed \Lb shape in the \LbTopKpipi decay are floated in the alternative fit to data.
The cross-feed backgrounds are described with one or two Crystal Ball functions with the tail and fraction parameters fixed from fits to simulated samples.
Ten thousand pseudoexperiments are generated using the alternative models with the same event yields determined in the fits to data.
The nominal model is then fitted to each generated sample and the asymmetry parameters are extracted.
As the bias observed is not significantly different from zero, the statistical uncertainty on the mean of the pulls is taken as the systematic uncertainty due to the model.

Further cross-checks are made to test the stability of the results with respect to different periods of data-taking, the different magnet polarities, the choice made in the selection of multiple candidates, and the effect of the trigger and selection criteria.
The results of these checks are all statistically compatible with the nominal results, and no systematic uncertainty is assigned.

\begin{table}[ht]
\centering
\small
\caption{\label{tab:Systematics}Sources of systematic uncertainty and their relative contributions to the total uncertainty.
Where present, the value in brackets shows the systematic uncertainty assigned to the measurement in specific phase-space regions.}
\begin{tabular}{llll}
\hline
Contribution			& \LbTopKpipi (\%)	& \LbTopKKK (\%)	& \XibTopKKpiWS (\%) \\
\hline
Experimental bias	& $\pm0.31 \; (\pm0.60)$	& $\pm0.31 \; (\pm0.60)$	&	$\pm0.31$  \\
\CT resolution		&	$\pm0.01$						&	$\pm0.05$						&	$\pm0.02$ \\
Fit model					&	$\pm0.03$						&	$\pm0.08$						&	$\pm0.19$	\\
\hline
Total							&	$\pm0.31 \; (\pm0.60)$	&	$\pm0.32 \; (\pm0.61)$	&	$\pm0.36$ \\
\hline
\end{tabular}

\end{table}

\section{Conclusions}
A search for $P$ and \CP violation is performed in four-body \Lb decays. Candidates are reconstructed in a data sample of $pp$ collisions collected with the \lhcb detector in 2011 and 2012, corresponding to an integrated luminosity of 3\invfb.
Samples of \LbTopKpipi, \LbTopKKK and \XibTopKKpiWS decays are reconstructed, yielding $19877\pm195$, $5297\pm83$ and $709\pm45$ signal candidates, respectively.
Two different measurements are made: one integrated over the phase space, and the other in specific phase-space regions.

No significant asymmetry is observed in the integrated measurements with a sensitivity of 0.8\% in \LbTopKpipi, 1.5\% in \LbTopKKK and 5.2\% in \XibTopKKpiWS decays, where the uncertainty is combined between statistical and systematic.
The measurements in regions of the phase space for \LbTopKpipi and \LbTopKKK decays are also all found to be consistent with conservation of both $P$ symmetry and \CP symmetry.

The $\Lb\to\proton\Km\chi_{c0}(1P)(\to\pi^+\pi^-)$ and $\Lb\to\proton\Km\chi_{c0}(1P)(\to K^+K^-)$ decays are observed for the first time. The yields and the corresponding statistical uncertainties are $336\pm25$ and $332\pm23$, respectively.

\section*{Acknowledgements}
%
%
\noindent We express our gratitude to our colleagues in the CERN
accelerator departments for the excellent performance of the LHC. We
thank the technical and administrative staff at the LHCb
institutes. We acknowledge support from CERN and from the national
agencies: CAPES, CNPq, FAPERJ and FINEP (Brazil); MOST and NSFC
(China); CNRS/IN2P3 (France); BMBF, DFG and MPG (Germany); INFN
(Italy); NWO (The Netherlands); MNiSW and NCN (Poland); MEN/IFA
(Romania); MinES and FASO (Russia); MinECo (Spain); SNSF and SER
(Switzerland); NASU (Ukraine); STFC (United Kingdom); NSF (USA).  We
acknowledge the computing resources that are provided by CERN, IN2P3
(France), KIT and DESY (Germany), INFN (Italy), SURF (The
Netherlands), PIC (Spain), GridPP (United Kingdom), RRCKI and Yandex
LLC (Russia), CSCS (Switzerland), IFIN-HH (Romania), CBPF (Brazil),
PL-GRID (Poland) and OSC (USA). We are indebted to the communities
behind the multiple open-source software packages on which we depend.
Individual groups or members have received support from AvH Foundation
(Germany), EPLANET, Marie Sk\l{}odowska-Curie Actions and ERC
(European Union), ANR, Labex P2IO and OCEVU, and R\'{e}gion
Auvergne-Rh\^{o}ne-Alpes (France), Key Research Program of Frontier
Sciences of CAS, CAS PIFI, and the Thousand Talents Program (China),
RFBR, RSF and Yandex LLC (Russia), GVA, XuntaGal and GENCAT (Spain),
Herchel Smith Fund, the Royal Society, the English-Speaking Union and
the Leverhulme Trust (United Kingdom).

\clearpage

{\noindent\bf\Large Appendices}

\appendix
\section{Observation of the $\boldsymbol{\Lb\to \chi_{c0}{(1P)} \proton \Km}$ decay}
\label{app:xic02}
\begin{figure}[htb]
\centering
\small
\includegraphics[width=0.45\textwidth]{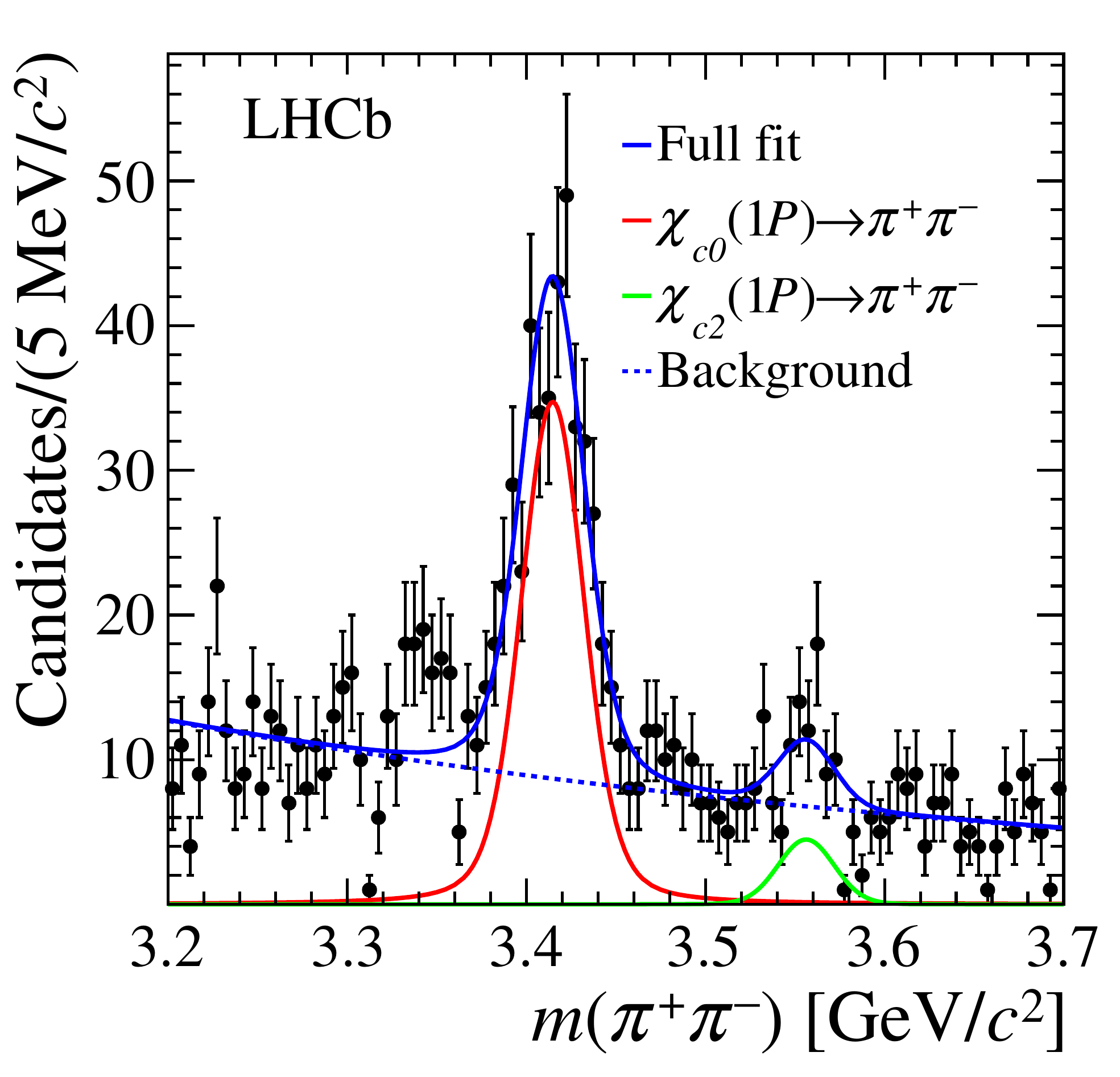}
\includegraphics[width=0.45\textwidth]{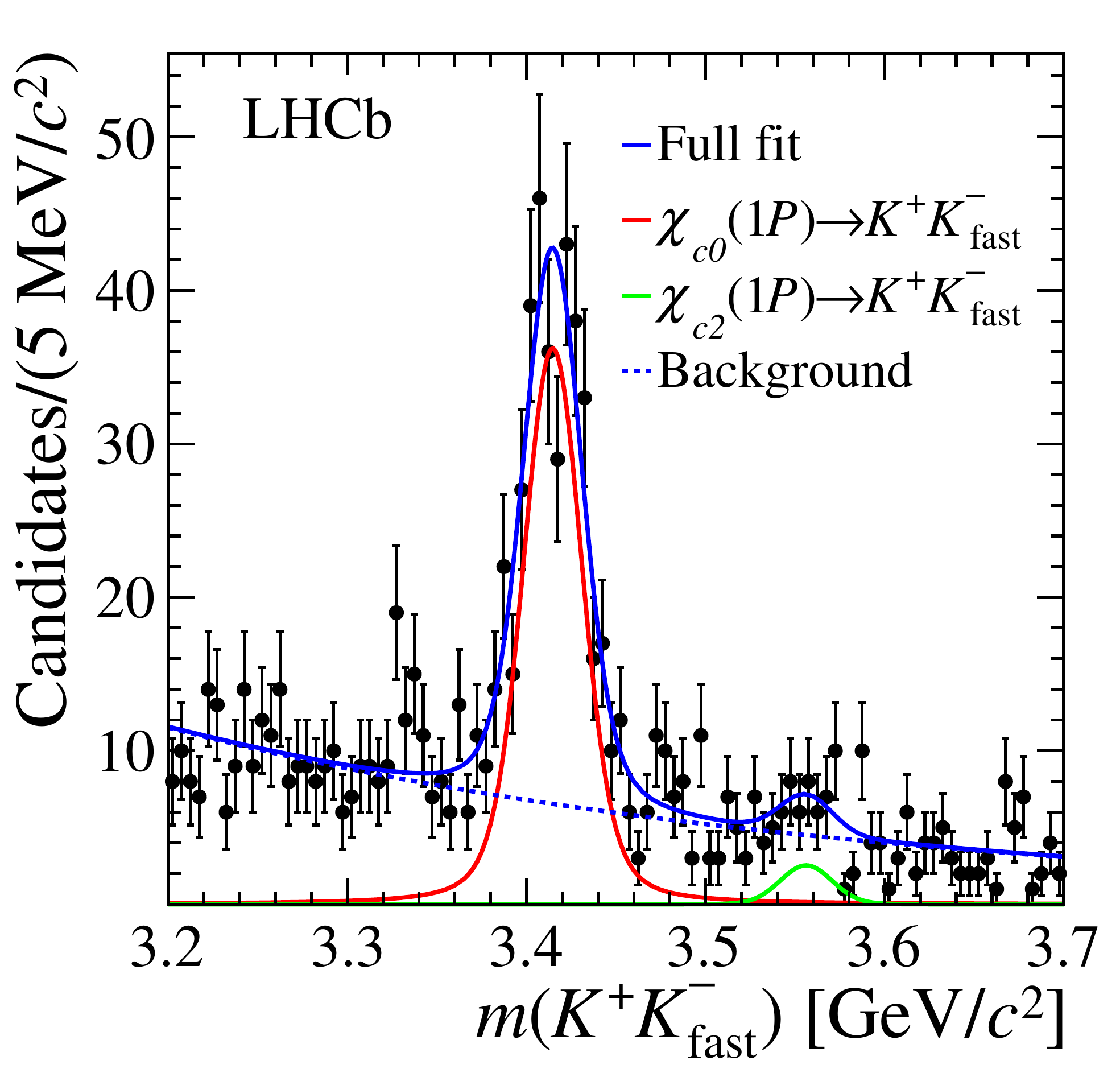}
	\caption{\small\label{fig:xic02}The left (right) plot shows the distribution of the $\pi^+\pi^-$ ($K^{+}K^{-}_{\textrm{fast}}$) reconstructed invariant mass for \Lb candidates selected within $\pm2\sigma$ of the \Lb mass peak. The results of the fit for different signal and background components are overlaid as described in the legend.}
\end{figure}

The $\pi^+\pi^-$ and $K^{+}K^{-}_{\textrm{fast}}$ invariant-mass distributions,
obtained by selecting \Lb candidates within a signal window of $\pm2\sigma$ with respect to the reconstructed \Lb mass peak, are shown in Fig.~\ref{fig:xic02}. The invariant mass distributions of the $\chi_{c0}{(1P)}$ and $\chi_{c2}{(1P)}$ signals are modelled by nonrelativistic Breit-Wigner functions convolved with a Gaussian function to account for the detector resolution.
The mean and width of the signal
Breit-Wigner functions are fixed to known values~\cite{PDG2014}, while the detector resolution, identical for the $\chi_{c0}{(1P)}$ and $\chi_{c2}{(1P)}$ signals,
is determined from the data.
The background, from random combinations of tracks and from \Lb decays that do not proceed via the $\chi_{c0}{(1P)}$ states, is modelled by an exponential function. An unbinned extended maximum likelihood fit is performed for $\pi^+\pi^-$ and $K^{+}K^{-}_{\textrm{fast}}$ invariant mass distributions. The signal yield for the $\Lb\to pK^-\chi_{c0}{(1P)}(\to\pi^+\pi^-)$ decay is $336\pm25$, and for $\Lb\to pK^-_{\textrm{slow}}\chi_{c0}{(1P)}(\to K^+K^-_{\textrm{fast}})$ decay is $332\pm23$,
where the uncertainty is statistical only. This represents the first
 observation of these decays. The signal yield and the statistical uncertainty for the $\Lb\to pK^-\chi_{c2}{(1P)}(\to\pi^+\pi^-)$ decay is $36\pm12$, and for $\Lb\to pK^-_{\textrm{slow}}\chi_{c2}{(1P)}(\to K^+K^-_{\textrm{fast}})$ decay is $19\pm9$.
%

\section{Measured asymmetries in regions of phase space}
\label{app:phase_space}
The definitions of the 14 (7) regions that form the binning scheme A (C) for the \LbTopKpipi (\LbTopKKK) decay are reported in Table~\ref{tab:PHSP_Def_SchemeA}~(\ref{tab:PHSP_Def_SchemeC}).
The measurements of \aCPTodd and \aPTodd in specific phase-space regions are reported in Table~\ref{tab:PHSP_Asym_SchemeAB}~(\ref{tab:PHSP_Asym_SchemeCD}).

\begin{table}[ht]
\centering
\small
	\caption{\label{tab:PHSP_Def_SchemeA}Definition of the 14 regions that form scheme A for the \LbTopKpipi decay. Bins $1-4$ focus on the region dominated by the $\Delta(1232)^{++}\to p\pip$ resonance. The other 10 bins are defined to study regions where $p\Km$ resonances are present on either side of the $f_0(980)\to\pip\pim$ or $\Kstarzb\to\Km\pip$ resonances. Further splitting depending on $|\Phi|$ is performed to reduce potential dilution of asymmetries, as suggested in Ref.~\cite{Durieux:2015zwa}. Masses are in units of \gevcc.
}
\begin{tabular}{cccccc}
\hline
Region	& $m(p\pip)$	& $m(p\Km)$	& $m(\pip\pim)$ & $m(\Km\pip)$	& $|\Phi|$\\
\hline
1		&	$(1.00,1.23)$	&		&			&  &	$(0,\frac{\pi}{2})$ \\
2		&	$(1.00,1.23)$	&		&			&  &	$(\frac{\pi}{2},\pi)$ \\
3		&	$(1.23,1.35)$	&		&			&  &	$(0,\frac{\pi}{2})$ \\
4		&	$(1.23,1.35)$	&		&			&  &	$(\frac{\pi}{2},\pi)$ \\
5		&	$(1.35,5.40)$	&	$(1.00,2.00)$	&	$(0.27,0.99)$	&  &	$(0,\frac{\pi}{2})$ \\
6		&	$(1.35,5.40)$	&	$(1.00,2.00)$	&	$(0.27,0.99)$	&  &	$(\frac{\pi}{2},\pi)$ \\
7		&	$(1.35,5.40)$	&	$(1.00,2.00)$	&	$(0.99,4.50)$	&  &	$(0,\frac{\pi}{2})$ \\
8		&	$(1.35,5.40)$	&	$(1.00,2.00)$	&	$(0.99,4.50)$	&  &	$(\frac{\pi}{2},\pi)$ \\
9		&	$(1.35,5.40)$	&	$(2.00,5.00)$	&	$(0.27,0.99)$	& $(0.63,0.89)$ &	$(0,\frac{\pi}{2})$ \\
10	&	$(1.35,5.40)$	&	$(2.00,5.00)$	&	$(0.27,0.99)$	& $(0.89,4.50)$ &	$(0,\frac{\pi}{2})$ \\
11		&	$(1.35,5.40)$	&	$(2.00,5.00)$	&	$(0.27,0.99)$	&  &	$(\frac{\pi}{2},\pi)$ \\
12		&	$(1.35,5.40)$	&	$(2.00,5.00)$	&	$(0.99,4.50)$	& $(0.63,0.89)$ &	$(0,\frac{\pi}{2})$ \\
13	&	$(1.35,5.40)$	&	$(2.00,5.00)$	&	$(0.99,4.50)$	& $(0.89,4.50)$ &	$(0,\frac{\pi}{2})$ \\
14	&	$(1.35,5.40)$	&	$(2.00,5.00)$	&	$(0.99,4.50)$	&  &	$(\frac{\pi}{2},\pi)$ \\
\hline
\end{tabular}

\end{table}
\begin{table}[ht]
\centering
\small
\caption{\label{tab:PHSP_Asym_SchemeAB}Measurements of \aPTodd and \aCPTodd in specific phase-space regions for the \LbTopKpipi decay. Each value is obtained through an independent fit to the candidates in the corresponding region of the phase space. Scheme A is defined in Table~\ref{tab:PHSP_Def_SchemeA} and divides the phase space according to dominant resonant contributions, while scheme B consists of twelve non-overlapping bins of width $\pi/12$ in $|\Phi|$.
}
\begin{tabular}{ccc}
\hline
Scheme A & \aPTodd (\%)	& \aCPTodd (\%)	 \\
\hline
1	& $-8.3\pm7.2\pm0.6$	& $-6.5\pm7.2\pm0.6$			\\
2	&	$-4.2\pm3.2\pm0.6$	& $-0.6\pm3.2\pm0.6$			\\
3	&	$\phantom{-}7.7\pm5.8\pm0.6$	& $-7.8\pm5.8\pm0.6$		\\
4	&	$-9.1\pm4.3\pm0.6$  & $-2.2\pm4.3\pm0.6$  	\\
5	&	$\phantom{-}2.1\pm4.9\pm0.6$  & $-0.4\pm4.9\pm0.6$		\\
6	&	$-2.3\pm5.0\pm0.6$  & $-0.5\pm5.0\pm0.6$   \\
7	&	$-1.0\pm3.0\pm0.6$  & $-0.1\pm3.0\pm0.6$   \\
8	&	$\phantom{-}4.2\pm3.7\pm0.6$  & $-1.3\pm3.7\pm0.6$ 	\\
9	&	$-1.4\pm5.1\pm0.6$  & $-0.3\pm5.1\pm0.6$	 \\
10&	$-0.8\pm2.7\pm0.6$  & $-3.0\pm2.7\pm0.6$	 \\
11&	$-0.9\pm2.5\pm0.6$  & $\phantom{-}3.5\pm2.5\pm0.6$  \\
12&	$-3.2\pm2.9\pm0.6$  & $-3.0\pm2.9\pm0.6$   \\
13&	$\phantom{-}0.7\pm1.5\pm0.6$  & $-0.9\pm1.5\pm0.6$		\\
14&	$\phantom{-}1.4\pm2.8\pm0.6$  & $-0.3\pm2.8\pm0.6$ 	\\
\hline
Scheme B & \aPTodd (\%)	& \aCPTodd (\%)	 \\
\hline
1	& $\phantom{-}0.6\pm2.1\pm0.6$	& $-3.5\pm2.1\pm0.6$		\\
2	&	$-0.3\pm2.2\pm0.6$	& $\phantom{-}1.8\pm2.2\pm0.6$		\\
3	&	$-2.8\pm2.5\pm0.6$	&	$-1.4\pm2.5\pm0.6$	\\
4	&	$\phantom{-}2.9\pm2.9\pm0.6$	&	$-4.7\pm2.9\pm0.6$	\\
5	&	$-3.3\pm3.0\pm0.6$	&	$-4.1\pm3.0\pm0.6$	\\
6	&	$\phantom{-}0.3\pm3.1\pm0.6$	&	$\phantom{-}1.4\pm3.1\pm0.6$ \\
7	&	$-2.6\pm3.3\pm0.6$ &	$\phantom{-}3.8\pm3.3\pm0.6$ \\
8	&	$\phantom{-}4.1\pm3.6\pm0.6$ & $-2.8\pm3.6\pm0.6$ 	\\
9	&	$-2.6\pm3.2\pm0.6$ &	$\phantom{-}1.7\pm3.2\pm0.6$ \\
10&	$\phantom{-}0.1\pm3.1\pm0.6$ &	$-0.7\pm3.1\pm0.6$ \\
11&	$-0.7\pm3.2\pm0.6$ &	$-2.2\pm3.2\pm0.6$ \\
12&	$-4.6\pm3.2\pm0.6$ &	$\phantom{-}1.3\pm3.2\pm0.6$ \\
\hline
\end{tabular}

\end{table}
\begin{table}[ht]
\centering
\small
\caption{\label{tab:PHSP_Def_SchemeC}Definition of the seven regions that form scheme C for the \LbTopKKK decay. The scheme is defined to study regions where $\proton K_{\rm slow}^-$ resonances are present ($1-3$) on either side of the $\Phi\to\Kp\Km$ resonances. Masses are in units of \gevcc.
}
\begin{tabular}{cccc}
\hline
Region	& $m(p K_{\rm slow}^-)$	& $m(\Kp K_{\rm slow}^-),m(\Kp K_{\rm fast}^-)$		& $|\Phi|$\\
\hline
1			& $(0.9,2.0)$	& $m(\Kp K_{\rm slow}^-)<1.02$~~ or $m(\Kp K_{\rm fast}^-)<1.02$  &	 \\
2			& $(0.9,2.0)$	& $m(\Kp K_{\rm slow}^-)>1.02$ and $m(\Kp K_{\rm fast}^-)>1.02$  & $(0,\frac{\pi}{2})$ 	 \\
3			& $(0.9,2.0)$	& $m(\Kp K_{\rm slow}^-)>1.02$ and $m(\Kp K_{\rm fast}^-)>1.02$  & $(\frac{\pi}{2},\pi)$ 	 \\
4			& $(2.0,4.0)$	& $m(\Kp K_{\rm slow}^-)<1.02$~~ or $m(\Kp K_{\rm fast}^-)<1.02$  & $(0,\frac{\pi}{2})$ 	 \\
5			& $(2.0,4.0)$	& $m(\Kp K_{\rm slow}^-)<1.02$~~ or $m(\Kp K_{\rm fast}^-)<1.02$  & $(\frac{\pi}{2},\pi)$ 	 \\
6			& $(2.0,4.0)$	& $m(\Kp K_{\rm slow}^-)>1.02$ and $m(\Kp K_{\rm fast}^-)>1.02$  & $(0,\frac{\pi}{2})$ 	 \\
7			& $(2.0,4.0)$	& $m(\Kp K_{\rm slow}^-)>1.02$ and $m(\Kp K_{\rm fast}^-)>1.02$  & $(\frac{\pi}{2},\pi)$ 	 \\
\hline
\end{tabular}

\end{table}
\begin{table}[ht]
\centering
\small
\caption{\label{tab:PHSP_Asym_SchemeCD}Measurements of \aPTodd and \aCPTodd in specific phase-space regions for the \LbTopKKK decay. Each value is obtained through an independent fit to the candidates in the corresponding region of the phase space. Scheme C is defined in Table~\ref{tab:PHSP_Def_SchemeC} and divides the phase space according to dominant resonant contributions, while scheme D consists of ten non-overlapping bins of width $\pi/10$ in $|\Phi|$.
}
\begin{tabular}{ccc}
\hline
Scheme C & \aPTodd (\%)	& \aCPTodd (\%)	 \\
\hline
1	& $\phantom{-}4.8\pm5.2\pm0.6$	 & $\phantom{-}6.0\pm5.2\pm0.6$			\\
2	&	$-2.8\pm2.5\pm0.6$	 & $\phantom{-}1.7\pm2.5\pm0.6$			\\
3	&	$\phantom{-}0.2\pm4.9\pm0.6$	 & $\phantom{-}0.2\pm4.9\pm0.6$		\\
4	&$-15.8\pm6.3\pm0.6$	 & $\phantom{-}0.4\pm6.3\pm0.6$  	\\
5	&	$\phantom{-}4.6\pm5.9\pm0.6$	 & $-2.5\pm5.9\pm0.6$		\\
6	&	$\phantom{-}2.8\pm3.7\pm0.6$   & $\phantom{-}0.9\pm3.7\pm0.6$   \\
7	&	$-2.7\pm3.4\pm0.6$   & $\phantom{-}1.5\pm3.4\pm0.6$   \\
\hline
Scheme D & \aPTodd (\%)	& \aCPTodd (\%)	 \\
\hline
1	& $-0.1\pm3.0\pm0.6$	& $-0.1\pm3.0\pm0.6$		\\
2	&	$-3.2\pm4.2\pm0.6$	& $\phantom{-}2.3\pm4.2\pm0.6$		\\
3	&	$-5.5\pm4.4\pm0.6$	&	$\phantom{-}1.7\pm4.4\pm0.6$	\\
4	&	$-2.0\pm5.1\pm0.6$	&	$\phantom{-}4.3\pm5.1\pm0.6$	\\
5	&	$-2.0\pm5.8\pm0.6$	&	$\phantom{-}1.5\pm5.8\pm0.6$	\\
6	&	$\phantom{-}3.1\pm5.5\pm0.6$	&	$-0.9\pm5.5\pm0.6$ \\
7	&	$\phantom{-}3.6\pm5.8\pm0.6$	&	$\phantom{-}2.5\pm5.8\pm0.6$ \\
8	&	$-6.6\pm5.9\pm0.6$	& $-0.5\pm5.9\pm0.6$ 	\\
9	&	$-6.6\pm5.6\pm0.6$	&	$-2.8\pm5.6\pm0.6$ \\
10&	$\phantom{-}6.2\pm5.7\pm0.6$	&	$\phantom{-}4.3\pm5.7\pm0.6$ \\
\hline
\end{tabular}

\end{table}

\section{Background-subtracted distributions in phase space}
\label{app:plot_phsp}
\begin{figure}[tb]
\centering
\small
\includegraphics[width=0.78\textwidth]{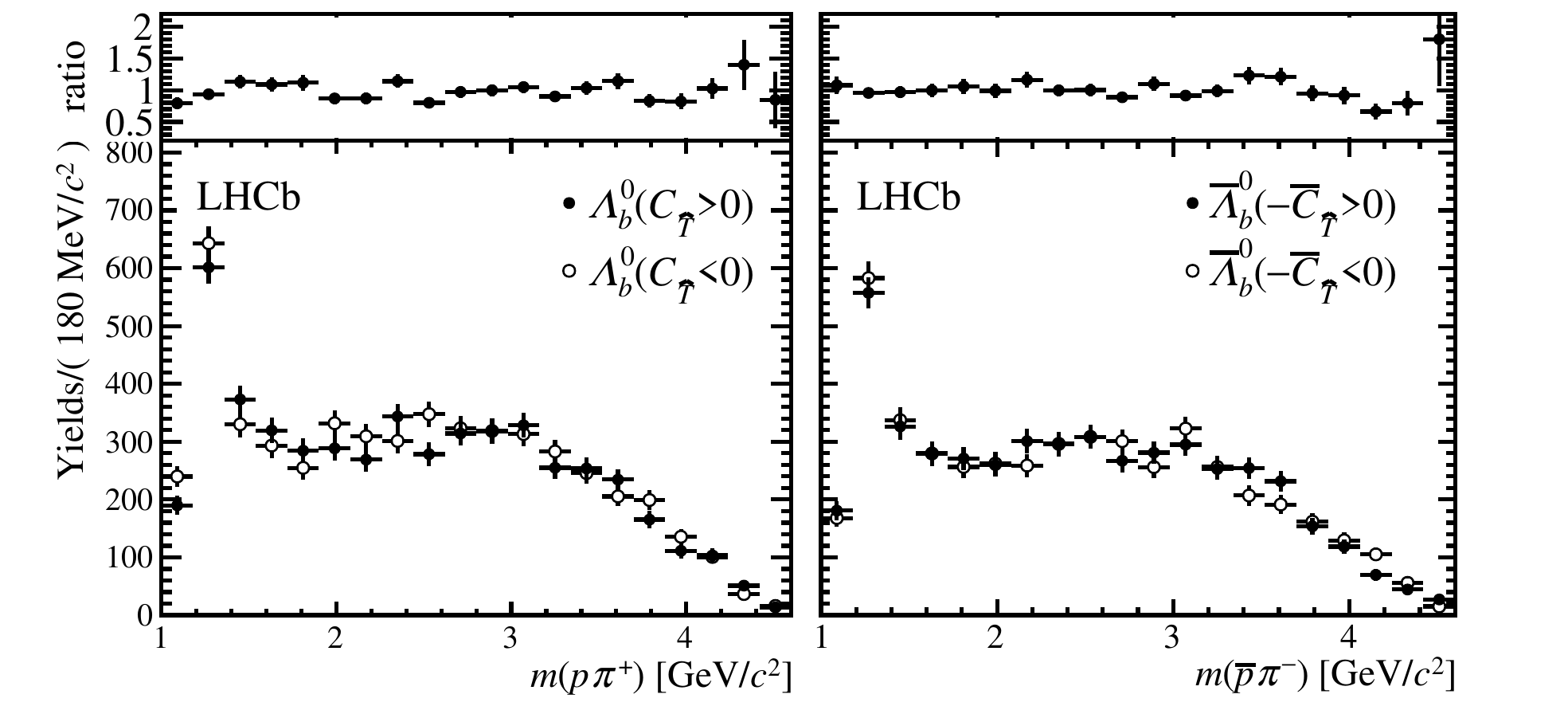}
\includegraphics[width=0.78\textwidth]{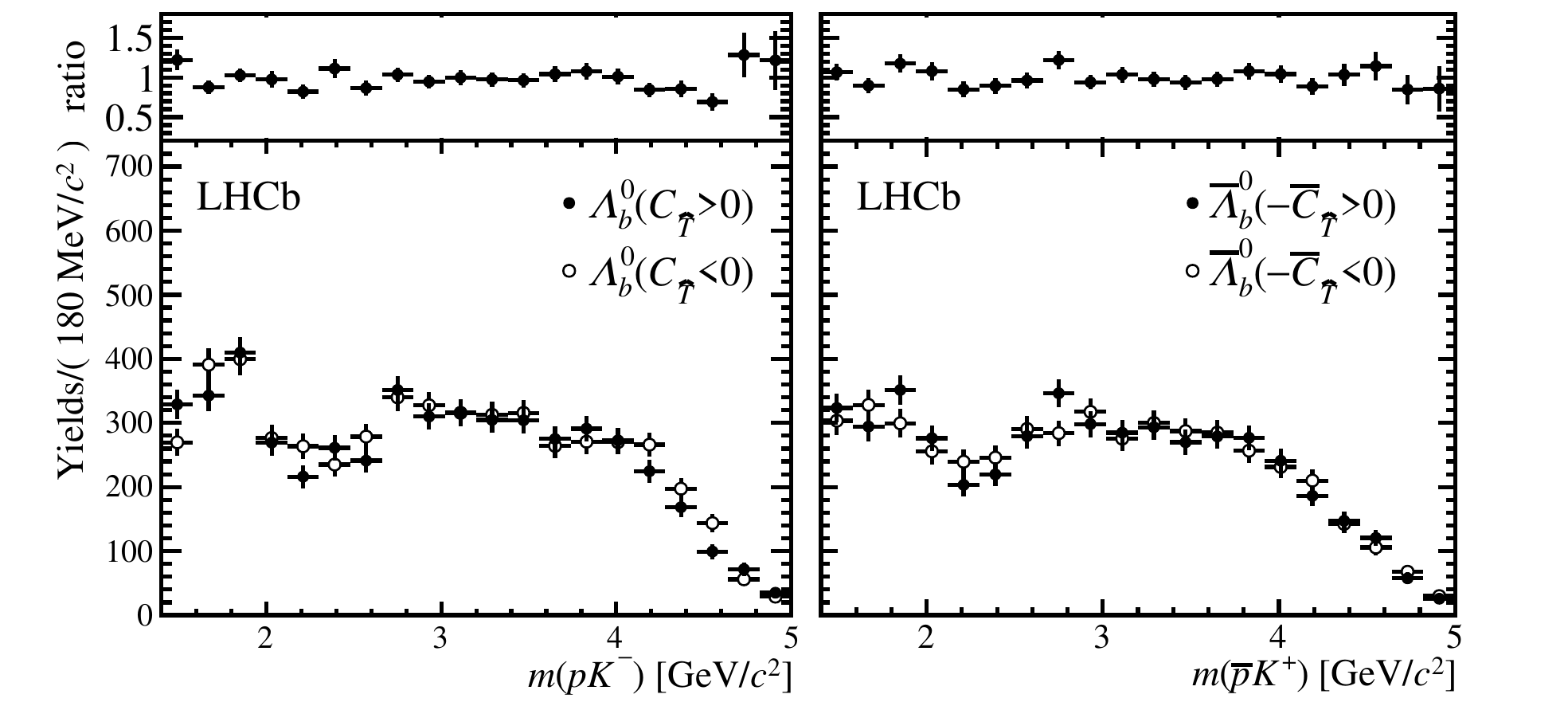}
\includegraphics[width=0.78\textwidth]{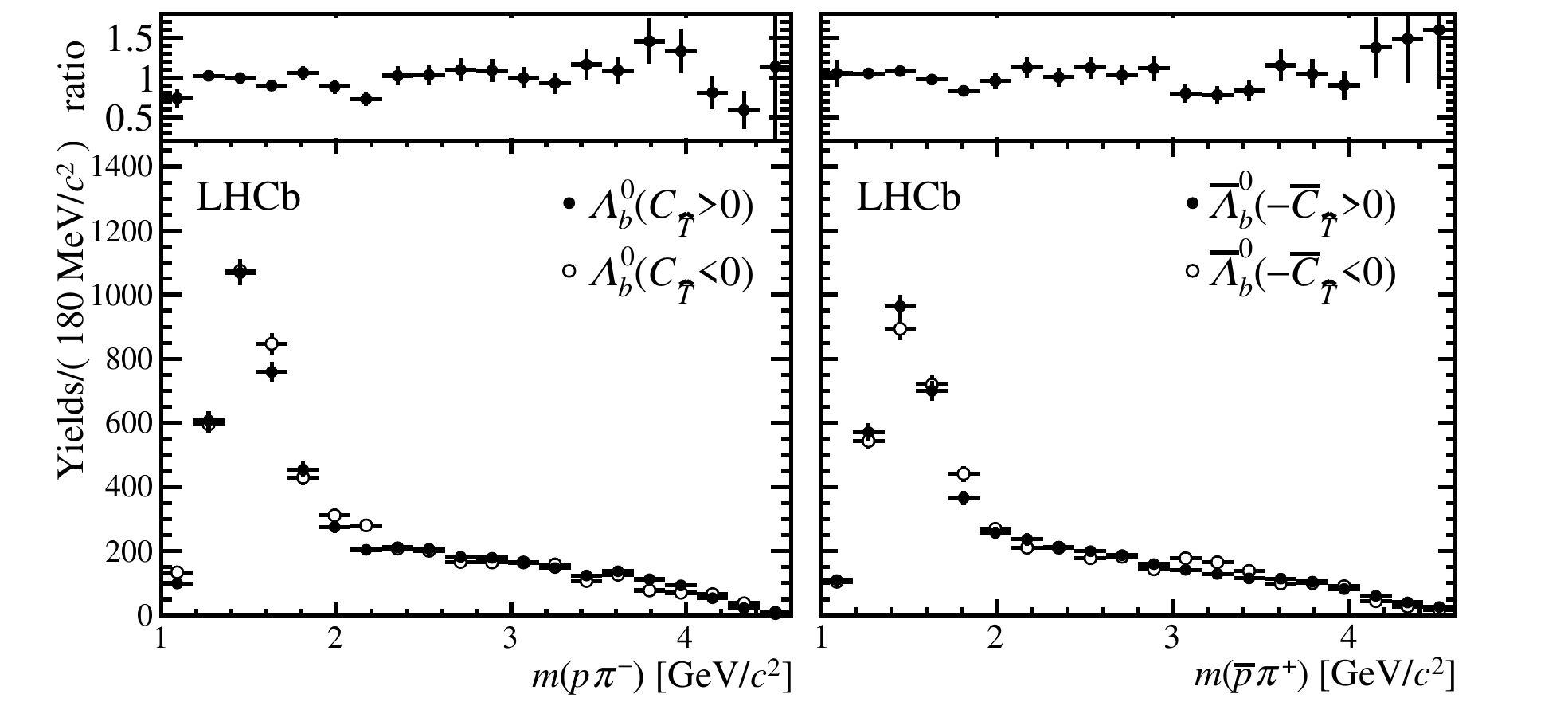}
\includegraphics[width=0.78\textwidth]{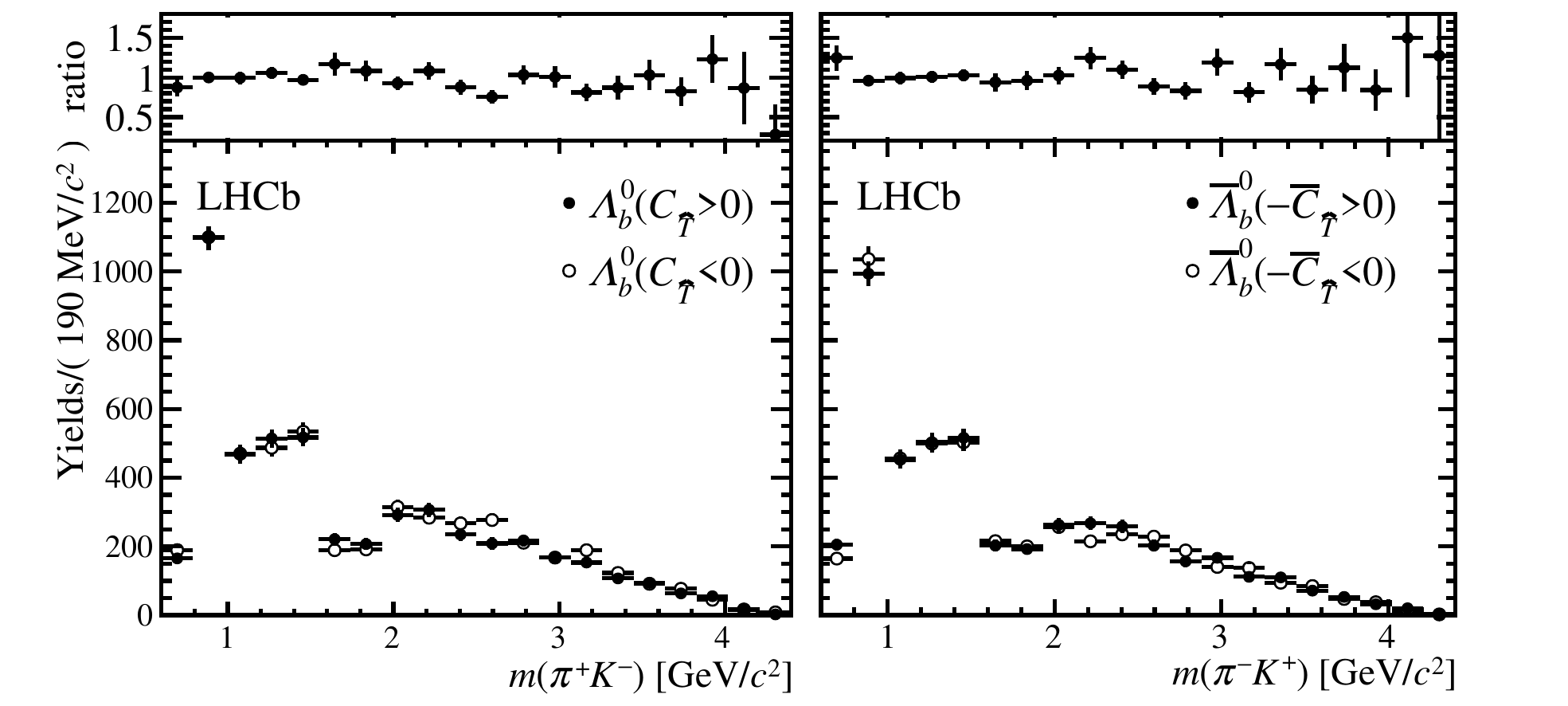}
\caption{\small\label{fig:app:phspLb2pK2pi}Background-subtracted distributions of \Lb (\Lbbar) candidates in different regions of phase space of the \LbTopKpipi decay for different values of \CT (\CTbar). The background subtraction is performed using the $sPlot$ technique~\cite{Pivk:2004ty}.}
\end{figure}

\begin{figure}[tb]
\centering
\small
\includegraphics[width=0.78\textwidth]{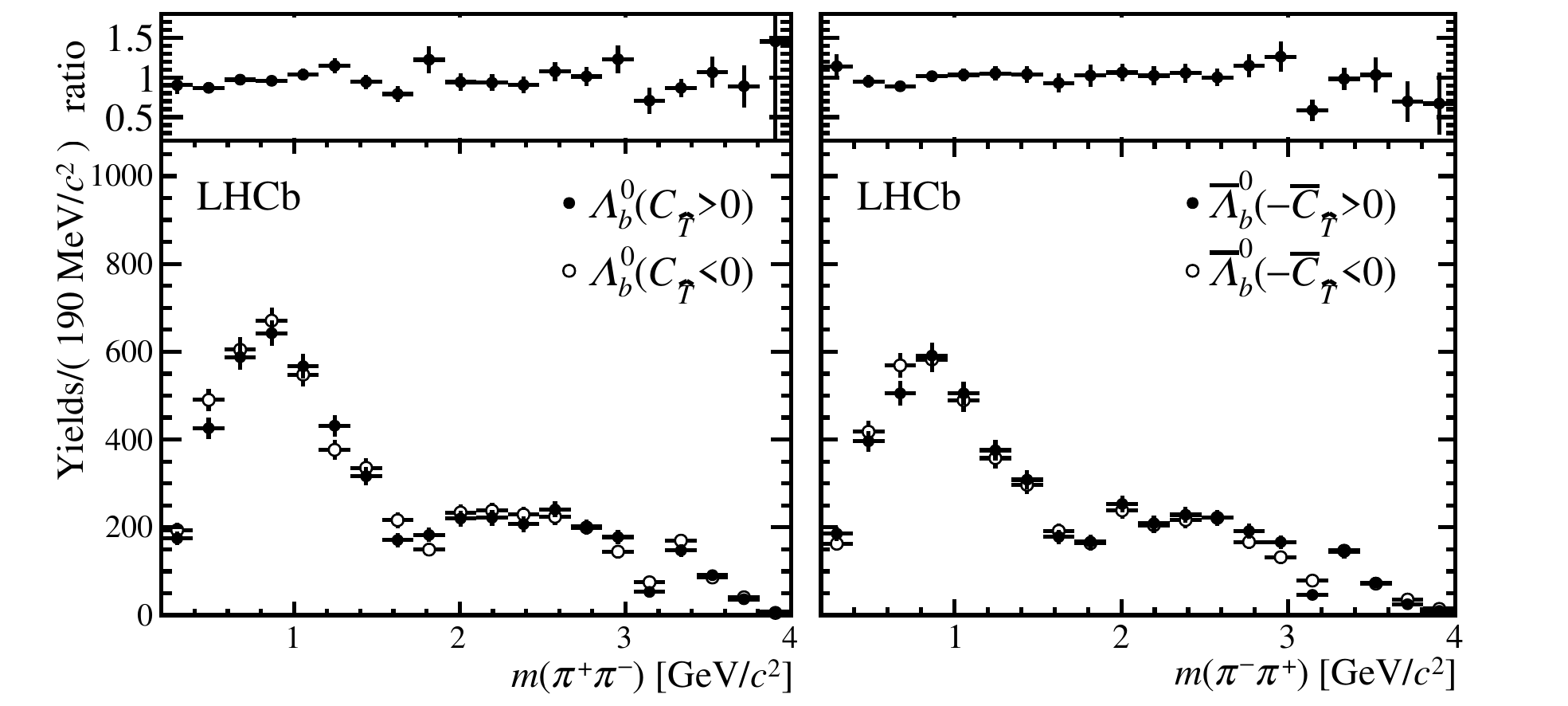}
\includegraphics[width=0.78\textwidth]{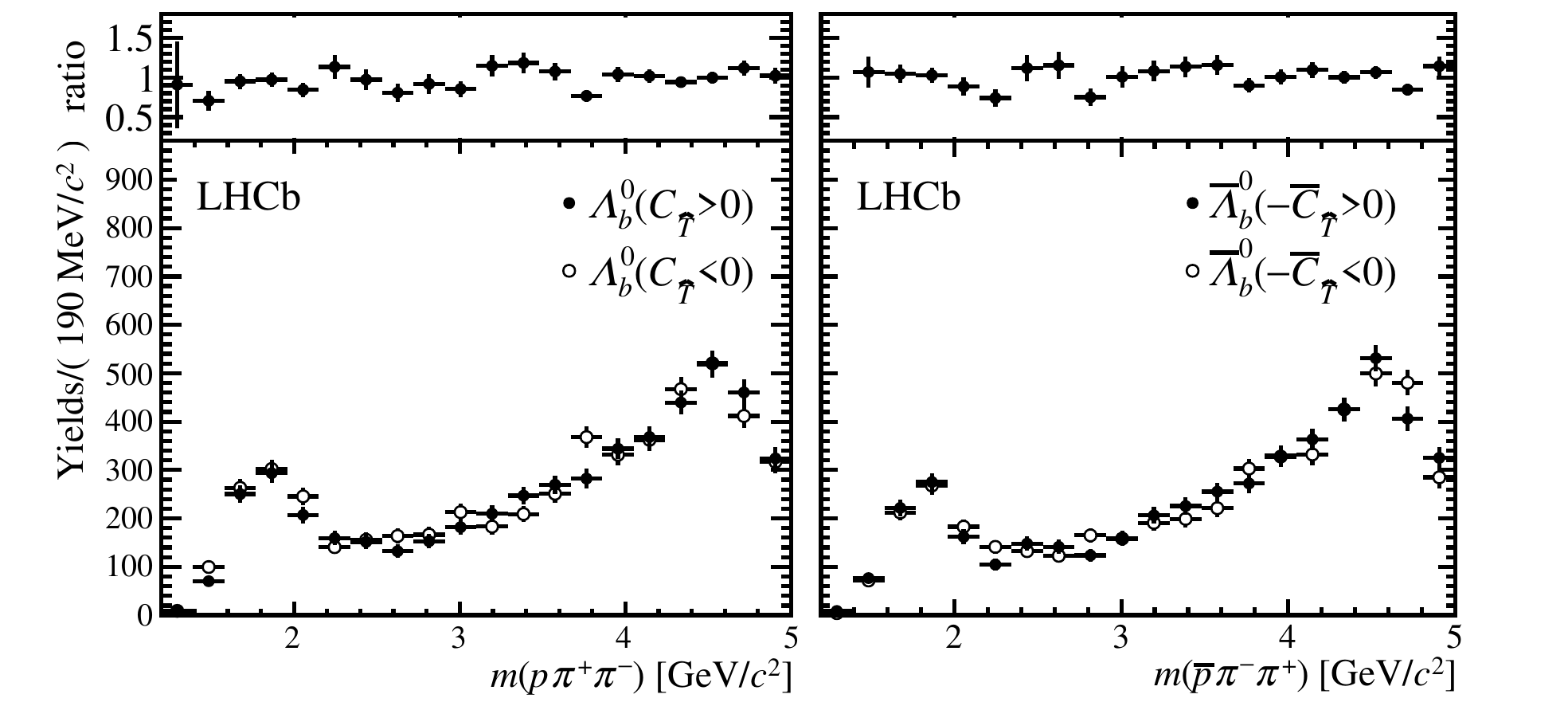}
\includegraphics[width=0.78\textwidth]{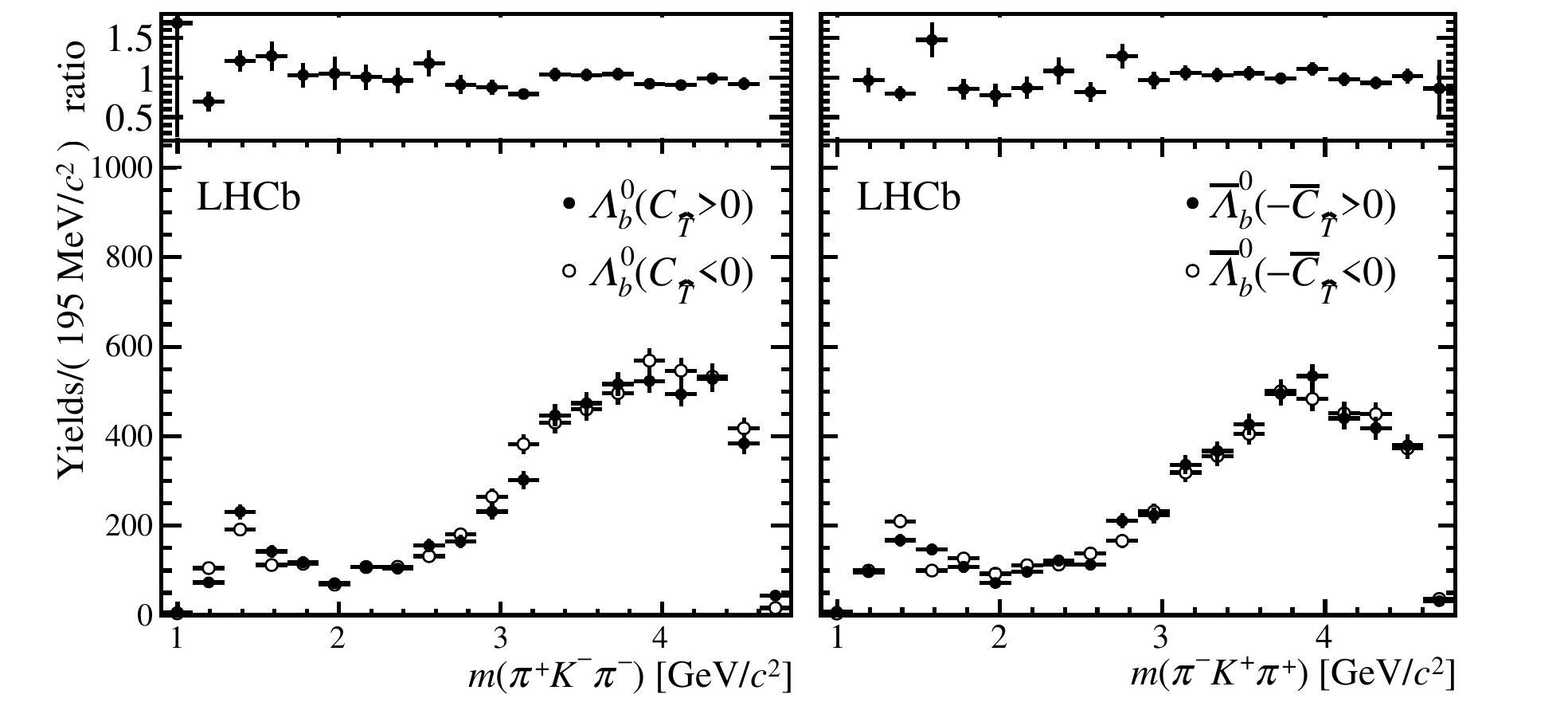}
\includegraphics[width=0.78\textwidth]{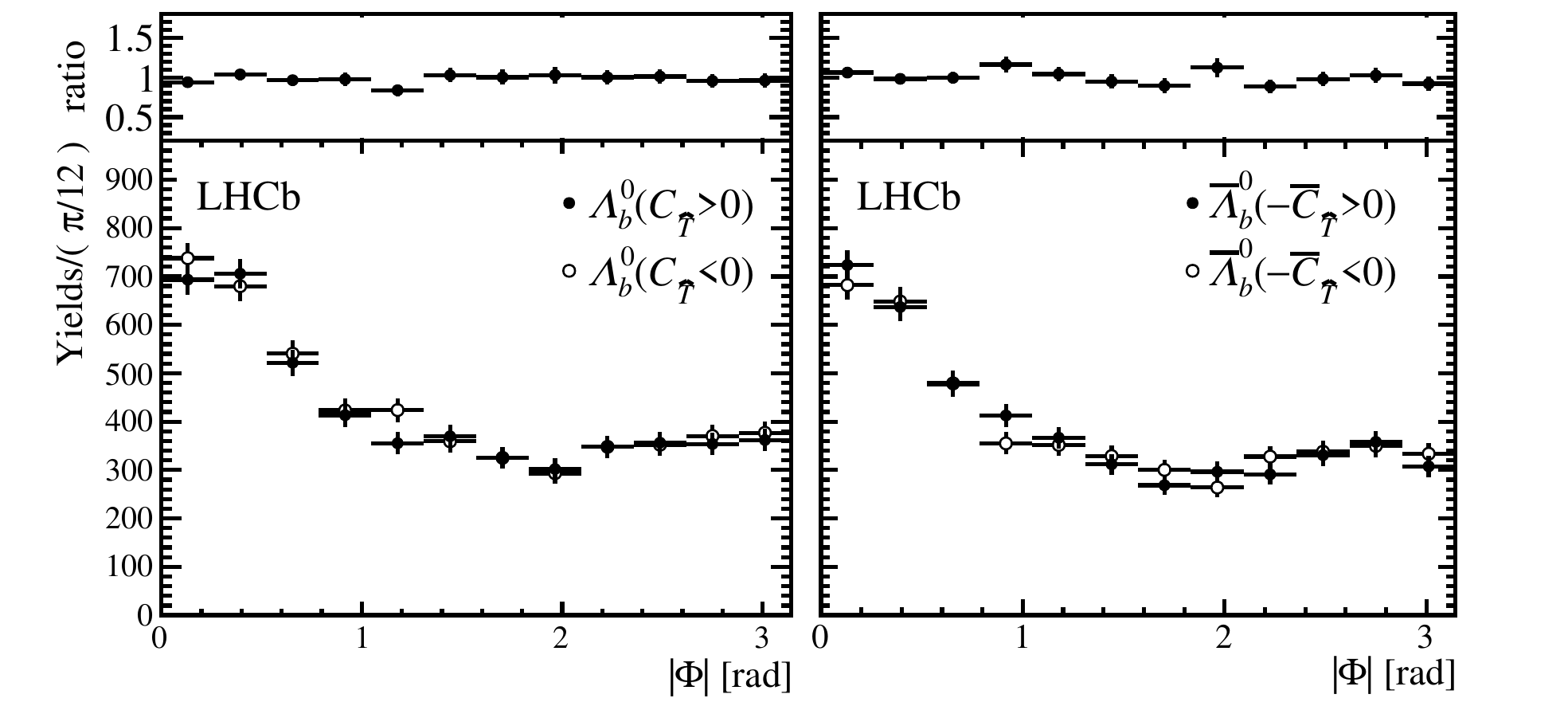}
\caption{\small\label{fig:app:phspLb2pK2pi-2}Background-subtracted distributions of \Lb (\Lbbar) candidates in different regions of phase space of the \LbTopKpipi decay for different values of \CT (\CTbar). The background subtraction is performed using the $sPlot$ technique~\cite{Pivk:2004ty}.}
\end{figure}

\begin{figure}[tb]
\centering
\small
\includegraphics[width=0.78\textwidth]{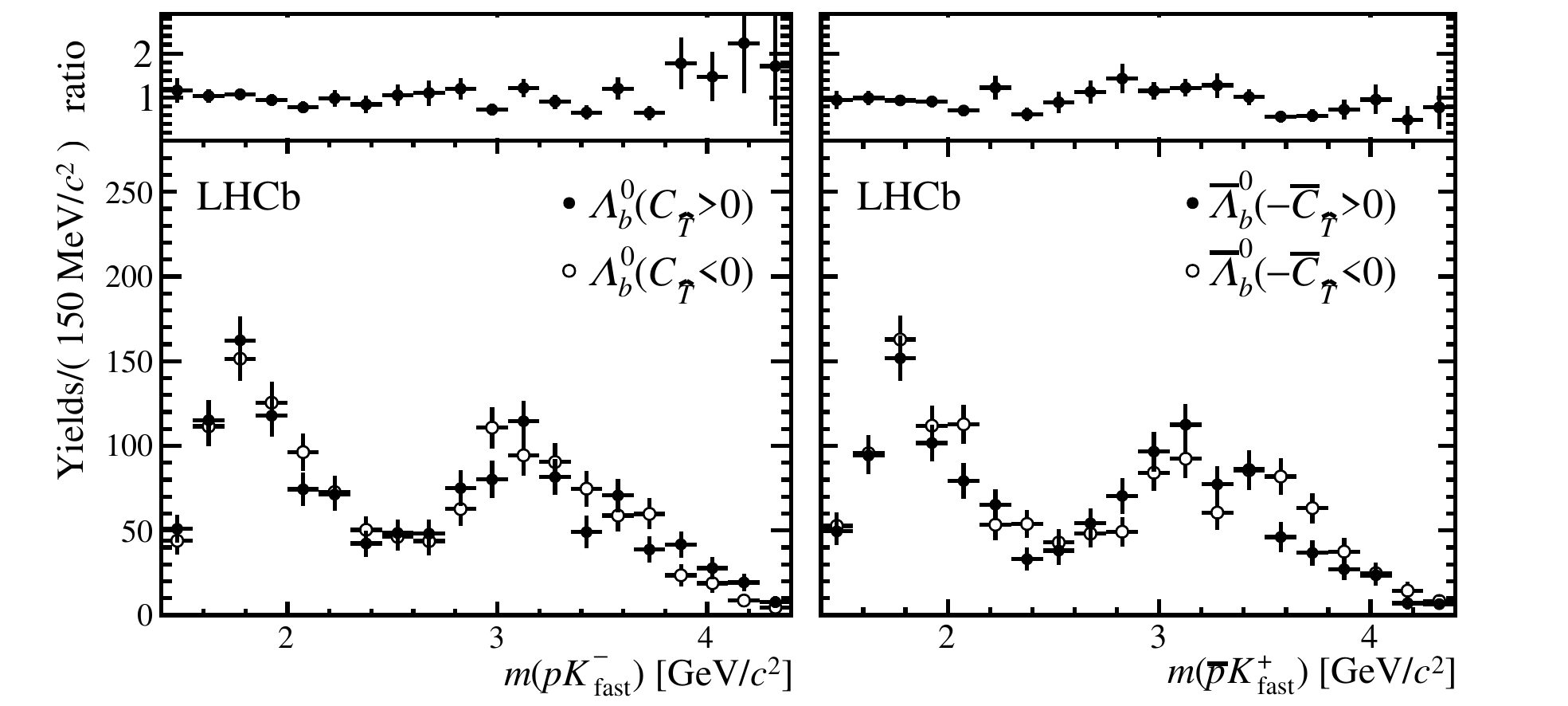}
\includegraphics[width=0.78\textwidth]{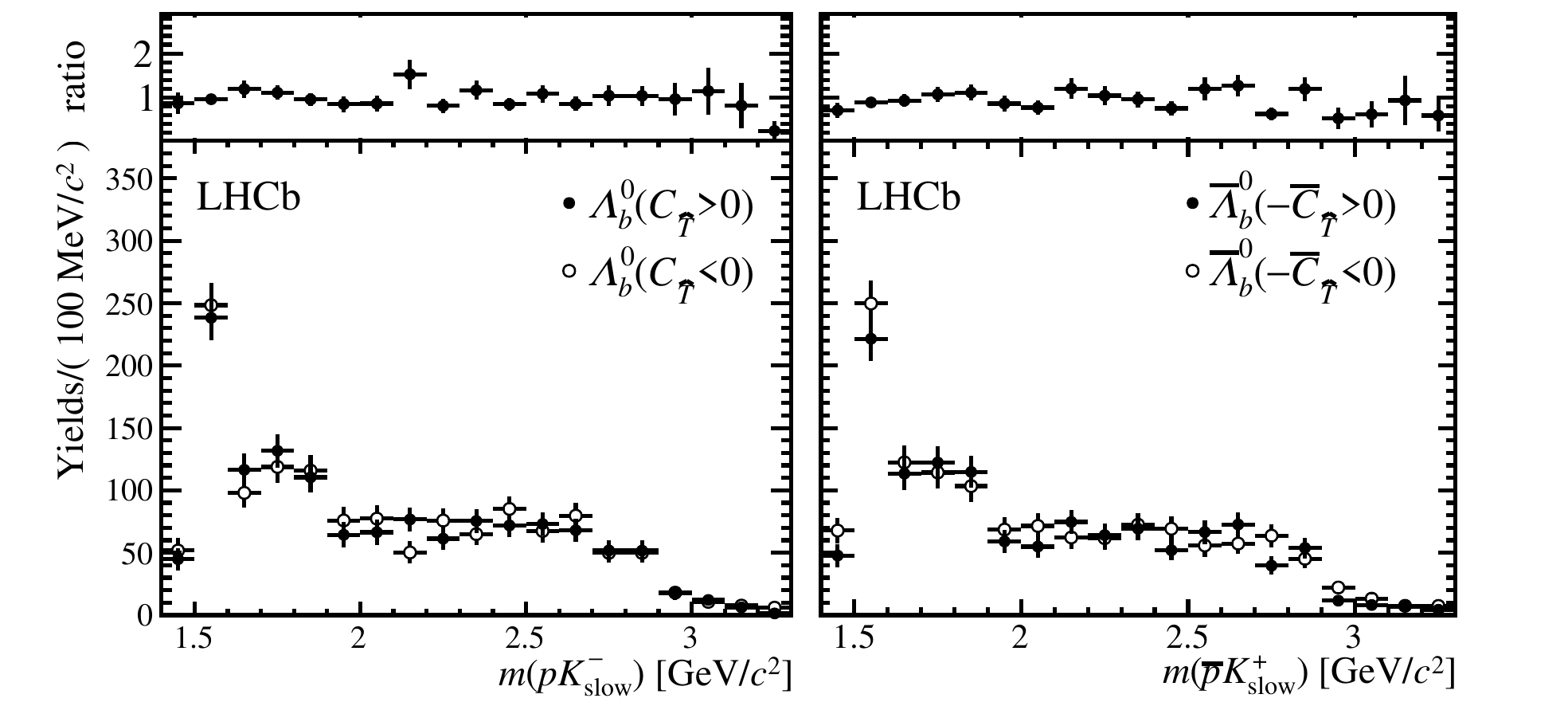}
\includegraphics[width=0.78\textwidth]{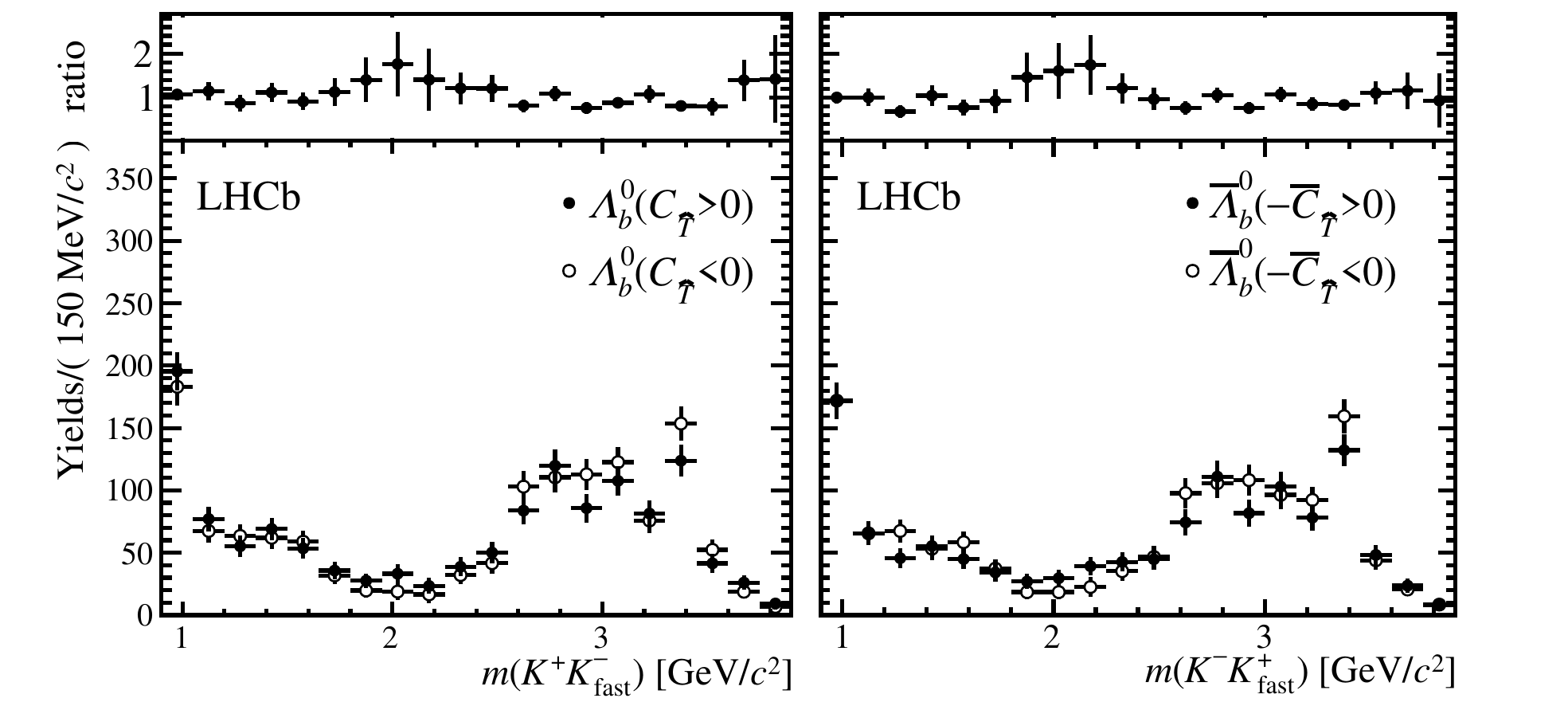}
\includegraphics[width=0.78\textwidth]{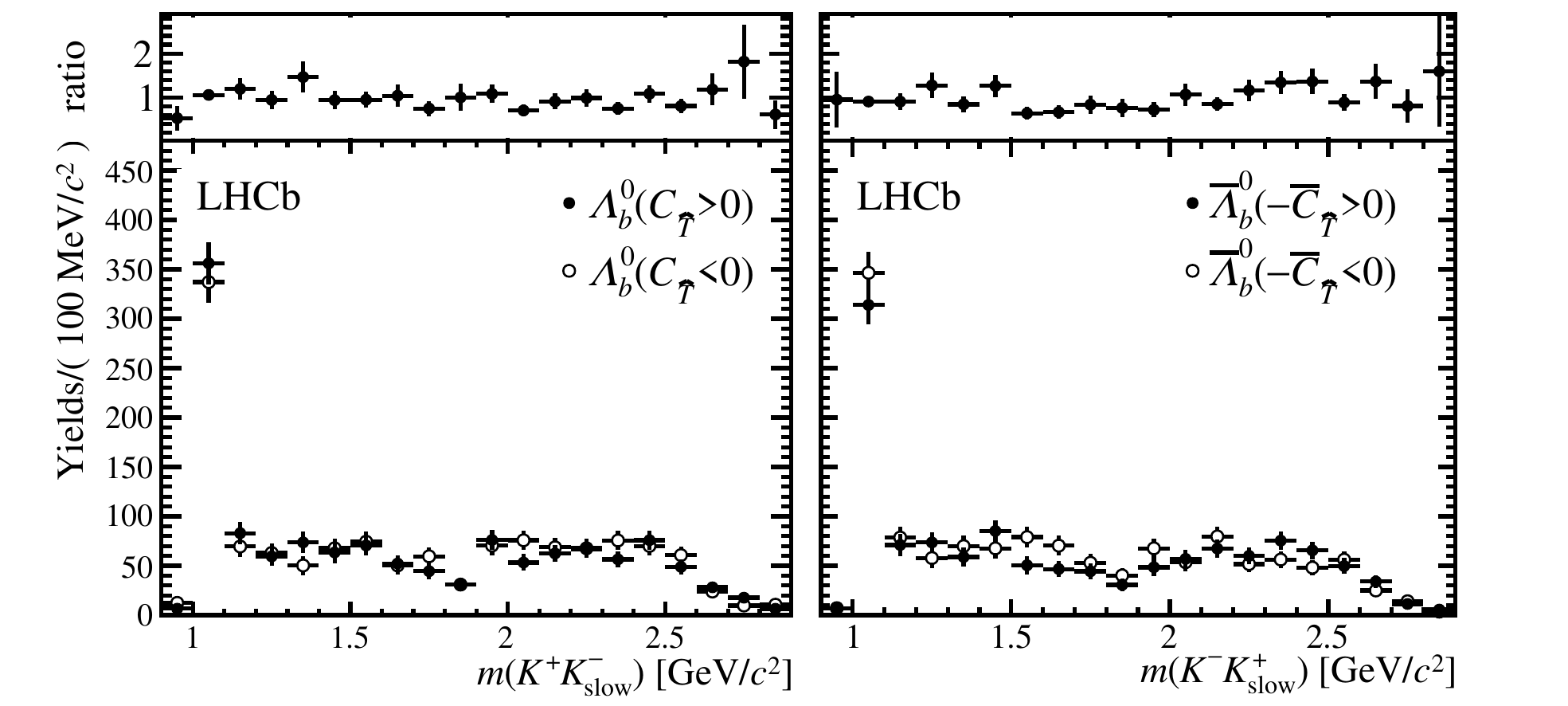}
\caption{\small\label{fig:app:phspLb2pKKK}Background-subtracted distributions of \Lb (\Lbbar) candidates in different regions of phase space of the \LbTopKKK decay for different values of \CT (\CTbar). The background subtraction is performed using the $sPlot$ technique~\cite{Pivk:2004ty}.}
\end{figure}

\begin{figure}[tb]
\centering
\small
\includegraphics[width=0.78\textwidth]{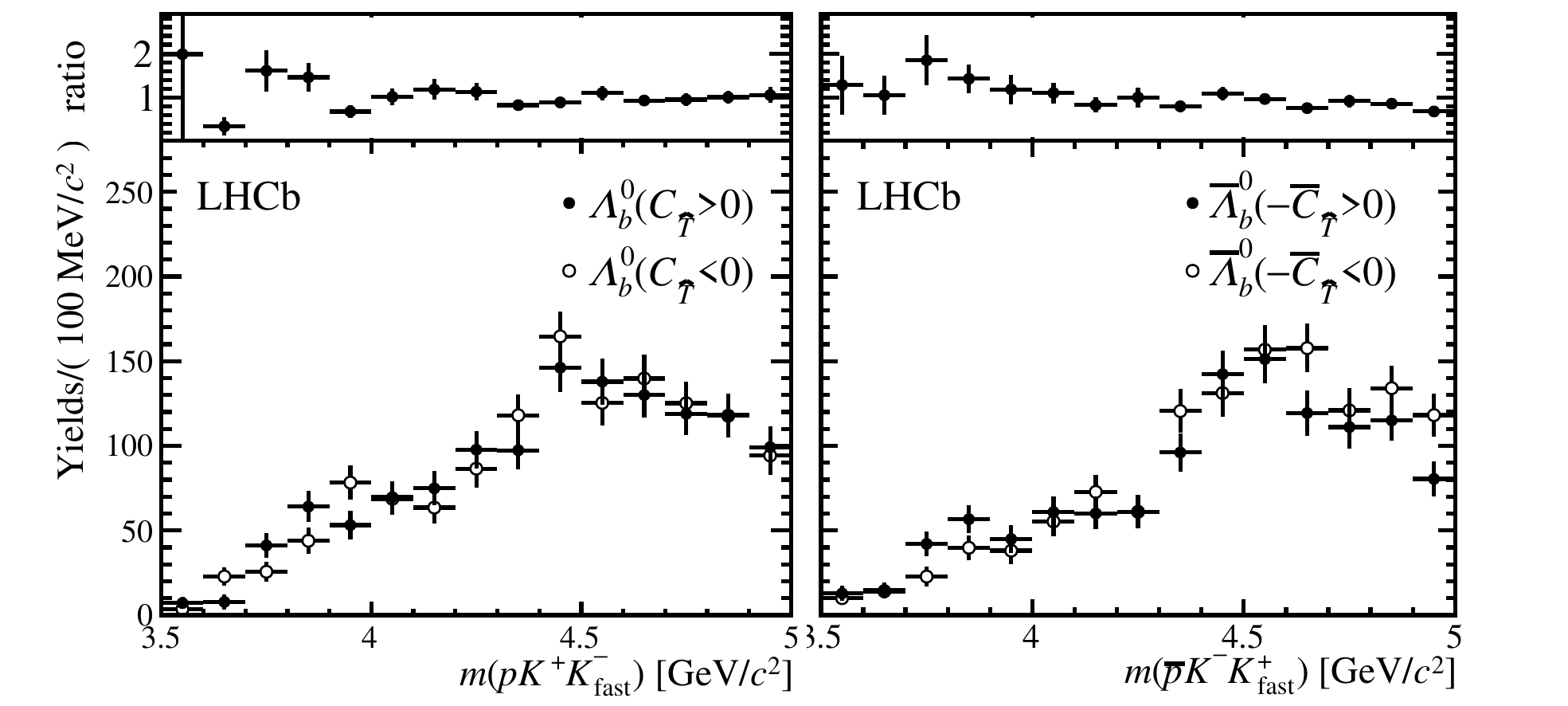}
\includegraphics[width=0.78\textwidth]{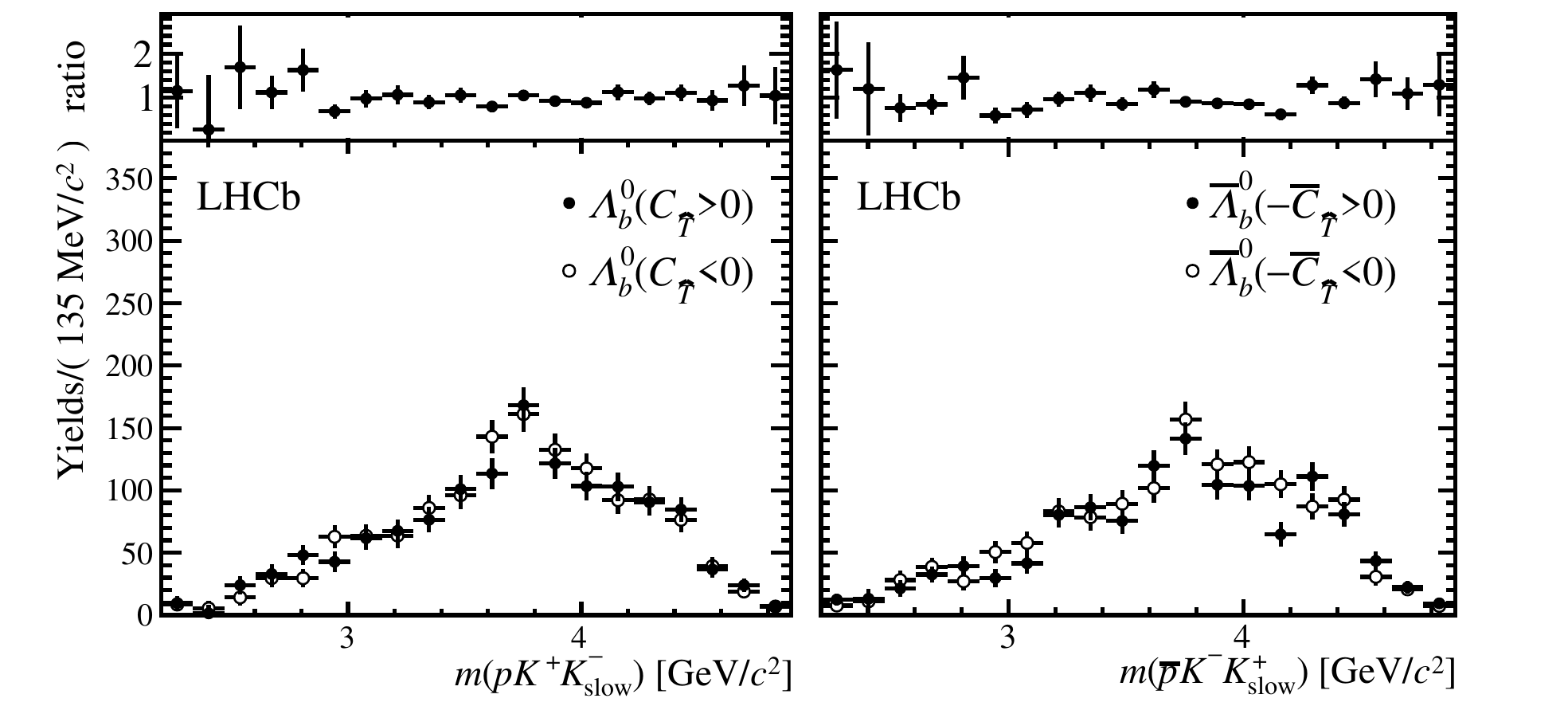}
\includegraphics[width=0.78\textwidth]{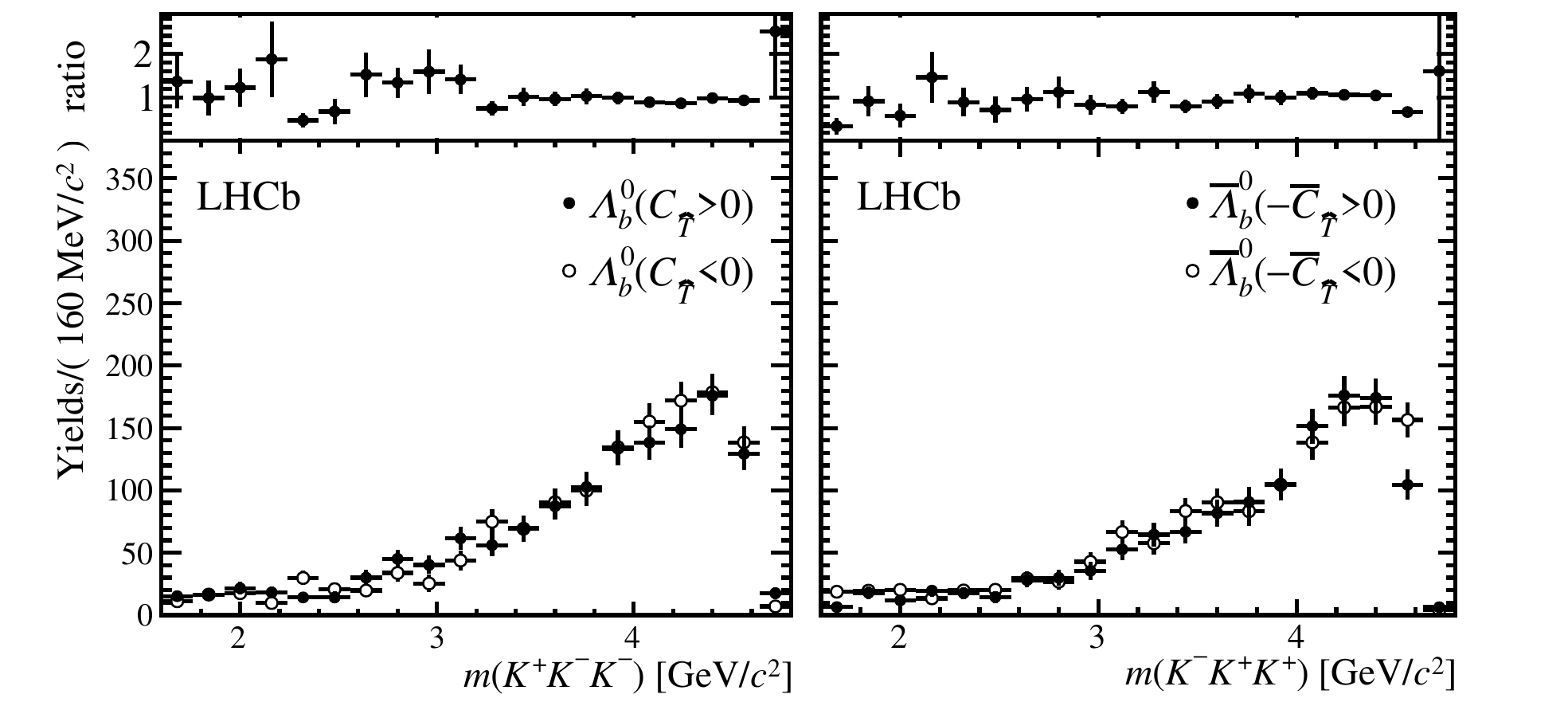}
\includegraphics[width=0.78\textwidth]{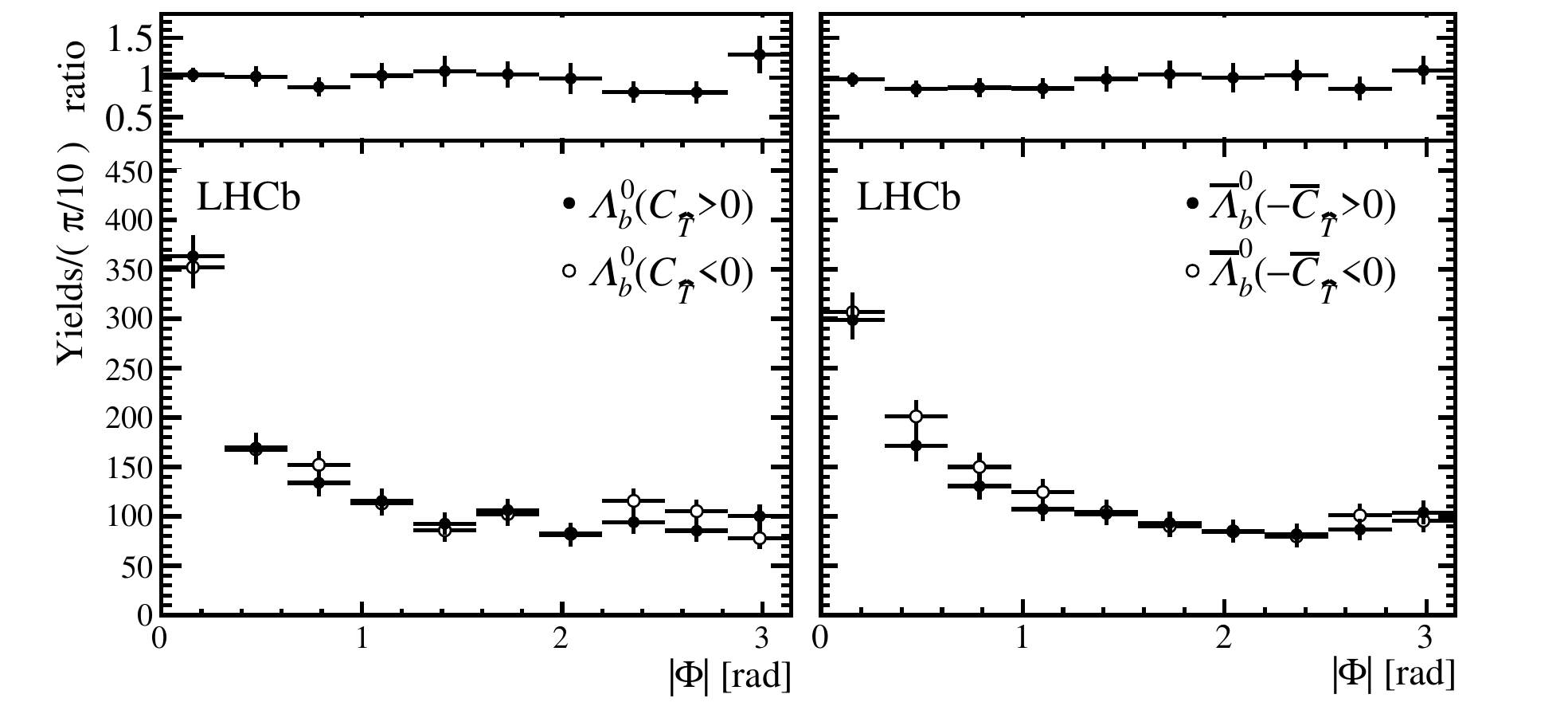}
\caption{\small\label{fig:app:phspLb2pKKK-2}Background-subtracted distributions of \Lb (\Lbbar) candidates in different regions of phase space of the \LbTopKKK decay for different values of \CT (\CTbar). The background subtraction is performed using the $sPlot$ technique~\cite{Pivk:2004ty}.}
\end{figure}

The background-subtracted distributions for \Lb(\Lbbar) with $\CT > 0$ and $\CT <0$ ( $-\CTbar > 0$ and $-\CTbar < 0$ ) in different regions of phase space of the \LbTopKpipi (\LbTopKKK) decay are shown in Figs.~\ref{fig:app:phspLb2pK2pi},~\ref{fig:app:phspLb2pK2pi-2}~(\ref{fig:app:phspLb2pKKK},~\ref{fig:app:phspLb2pKKK-2}). The distributions are made using the $sPlot$ technique~\cite{Pivk:2004ty}.

\clearpage


\clearpage

\addcontentsline{toc}{section}{References}
\setboolean{inbibliography}{true}
\bibliographystyle{LHCb}
\bibliography{main,this_paper,LHCb-PAPER,LHCb-CONF,LHCb-DP,LHCb-TDR}

\newpage

\centerline{\large\bf LHCb collaboration}
\begin{flushleft}
\small
R.~Aaij$^{40}$,
B.~Adeva$^{39}$,
M.~Adinolfi$^{48}$,
Z.~Ajaltouni$^{5}$,
S.~Akar$^{59}$,
J.~Albrecht$^{10}$,
F.~Alessio$^{40}$,
M.~Alexander$^{53}$,
A.~Alfonso~Albero$^{38}$,
S.~Ali$^{43}$
% ...

{\footnotesize \it
% ...
$ ^{5}$Clermont Universit{\'e}, Universit{\'e} Blaise Pascal, CNRS/IN2P3, LPC, Clermont-Ferrand, France\\
% ...
$ ^{10}$Fakult{\"a}t Physik, Technische Universit{\"a}t Dortmund, Dortmund, Germany\\
% ...
$ ^{38}$ICCUB, Universitat de Barcelona, Barcelona, Spain\\
$ ^{39}$Instituto Galego de F{\'\i}sica de Altas Enerx{\'\i}as (IGFAE), Universidade de Santiago de Compostela, Santiago de Compostela, Spain\\
$ ^{40}$European Organization for Nuclear Research (CERN), Geneva, Switzerland\\
% ...
$ ^{43}$Nikhef National Institute for Subatomic Physics, Amsterdam, The Netherlands\\
% ...
$ ^{48}$H.H. Wills Physics Laboratory, University of Bristol, Bristol, United Kingdom\\
% ...
$ ^{53}$School of Physics and Astronomy, University of Glasgow, Glasgow, United Kingdom\\
% ...
$ ^{59}$University of Cincinnati, Cincinnati, OH, United States\\
% ...
}
\end{flushleft}



\end{document}